\documentclass[twocolumn,showpacs,aps,epsfig,nofootinbib]{revtex4}

\usepackage[pdftex]{graphicx}
\usepackage{epstopdf}
\usepackage{latexsym}
\usepackage{amssymb}
\usepackage[normalem]{ulem} 

\usepackage[center]{subfigure}

\begin{document}

 \newcommand{\bq}{\begin{equation}}
 \newcommand{\eq}{\end{equation}}
 \newcommand{\bqn}{\begin{eqnarray}}
 \newcommand{\eqn}{\end{eqnarray}}
 \newcommand{\nb}{\nonumber}
 \newcommand{\lb}{\label}
\newcommand{\PRL}{Phys. Rev. Lett.}
\newcommand{\PL}{Phys. Lett.}
\newcommand{\PR}{Phys. Rev.}
\newcommand{\CQG}{Class. Quantum Grav.}

\begin{flushright}

{\mbox{\hspace{10cm}}} USTC-ICTS-11-04

\end{flushright}

\title{ Black holes and global structures of spherical  spacetimes   in Horava-Lifshitz theory } 

\author{Jared Greenwald ${}^{a}$}
\email{Jared\_Greenwald@baylor.edu}

\author{Jonatan Lenells ${}^{b}$}
\email{Jonatan\_Lenells@baylor.edu}

\author{ J. X.  Lu ${}^{c}$}
\email{jxlu@ustc.edu.cn}

\author{V. H. Satheeshkumar ${}^{a}$}
\email{VH\_Satheeshkumar@baylor.edu}

\author{Anzhong Wang ${}^{a,d}$}
\email{Anzhong\_Wang@baylor.edu}

\affiliation{${}^{a}$ GCAP-CASPER, Physics Department, Baylor
University, Waco, TX 76798-7316, USA\\
${}^{b}$ Mathematics  Department, Baylor
University, Waco, TX 76798-7328, USA\\
${}^{c}$  Interdisciplinary Center for Theoretical Study, 
University of Science and Technology of China, Hefei, Anhui 230026, China \\
${}^{d}$ Department of Physics, Zhejiang University of
Technology, Hangzhou 310032,  China}

\date{\today}

\begin{abstract}

We  systematically study black holes in the Horava-Lifshitz (HL) theory by following  the kinematic approach, in which a 
horizon is defined as the surface at which  massless test particles are infinitely redshifted. Because of the nonrelativistic  
dispersion relations, the speed of light is unlimited, and test particles do not follow geodesics. As a    result, there are 
significant differences in causal structures and black holes between general relativity (GR) and the HL theory. In particular, 
the horizon radii generically depend on the energies of test particles.  Applying them to the spherical static vacuum solutions 
found recently in  the nonrelativistic general covariant theory of gravity,  we find that, for test particles with sufficiently high 
energy, the radius of the horizon can be made as small as desired, although the singularities can be seen in principle only 
by observers with  infinitely high energy. In these studies, we pay  particular attention to the global structure of the solutions, 
and find that, because of the foliation-preserving-diffeomorphism symmetry, ${\mbox{Diff}}(M,{\cal{F}})$, they are quite 
different from the corresponding ones given in GR, even though the solutions are the same. In particular,  the 
${\mbox{Diff}}(M,{\cal{F}})$ does not allow Penrose diagrams. Among the vacuum solutions,  some   give rise to the
structure of the Einstein-Rosen bridge, in which two asymptotically flat regions are connected by a throat with a finite 
non-zero radius.  We also study  slowly rotating solutions in such a setup, and obtain all the solutions characterized by an
arbitrary function $A_{0}(r)$. The  case $A_{0} = 0$  reduces to  the slowly rotating Kerr solution  obtained  in GR.

\end{abstract}


\maketitle

\section{Introduction}
\renewcommand{\theequation}{1.\arabic{equation}} \setcounter{equation}{0}

Horava-Lifshitz (HL) theory, proposed recently by  Horava \cite{Horava}, and  motivated by the Lifshitz 
theory of a scalar field with anisotropic scalings  \cite{Lifshitz}, 
\bq
\lb{1.1}
{\bf x} \rightarrow \ell {\bf x}, \;\;\;  t \rightarrow \ell^{z} t,\; (z \not = 1), 
\eq
has attracted 
lot of attention, due to its several remarkable
features. In particular, the effective speed of light in this  theory diverges in the ultraviolet (UV), which could potentially resolve the 
horizon problem without invoking inflation \cite{KKa}. The  spatial curvature is enhanced by higher-order 
curvature terms, and this opens a new approach to investigating both the flatness problem and 
 bouncing universes \cite{Calcagni,brand,WWa}. In addition,  in the super-horizon region scale-invariant  
curvature perturbations can be produced without inflation \cite{Muka,KKa,Piao,YKN,WM}. The perturbations 
become adiabatic during slow-roll inflation driven by a single field,  and the comoving curvature perturbation 
is constant  \cite{WWM}.  For more detail, we refer readers to \cite{Mukc,Sotiriou,Padilla,Hreview,Visserb}.

Despite all  these remarkable features, the theory is plagued with three major problems, {\em   ghosts,  strong 
coupling and instability}.  Although they are different, their origins are  the same:  
the breaking of the general covariance  \cite{BKW}. The preferred time that breaks general covariance leads to a reduced
set of diffeomorphisms, 
\bq
\lb{1.2}
\tilde{t} = t - f(t),\; \;\; \tilde{x}^{i}  =  {x}^{i}  - \zeta^{i}(t, {\bf x}),
\eq
often denoted by Diff($M, \; {\cal{F}}$). As a result, a spin-0 graviton appears. This mode is potentially dangerous and may cause
the instability, ghost and  strong coupling problems, which   could prevent the recovery of
general relativity (GR) in the IR  \cite{Mukc,Sotiriou,Padilla,Hreview,Visserb}. 

To resolve these problems, various modifications have been proposed.  But, so far there are
{ only} two  that seem to have the potential to solve these problems: One  is due to Blas, Pujolas, and  Sibiryakov (BPS) 
\cite{BPSc}, who introduced a vector field 
$$
a_{i} = \partial_{i}\ln(N), 
$$
where $N$ denotes the lapse function \footnote{It is clear that
the BPS model works only for the  $N = N(t, x)$ case, in which the projectability condition $N = N(t)$ is broken. Otherwise, 
the vector field $a_{i}$ will vanish identically.  However, violation of  the projectability condition often leads to the inconsistency 
problem \cite{LP}. But, as shown in \cite{Kluson}, this
is not the case in the BPS model.  The inclusion of the vector field $a_{i}$ gives rise  to   a proliferation of independent  coupling constants
\cite{KP},  which  could potentially limit   the predictive powers of the theory.}. The other is due to Horava and 
Melby-Thompson (HMT)  \cite{HMT}, in which the projectability condition,
\bq
\lb{PorjC}
N = N(t),
\eq
  was assumed. In the HMT setup, 
 the foliation-preserving-diffeomorphisms Diff($M, \; {\cal{F}}$) are extended
to include  a local $U(1)$ symmetry, so that the total symmetry of the theory is enlarged to,
\bq
\lb{1.3}
 U(1) \ltimes {\mbox{Diff}}(M, \; {\cal{F}}). 
 \eq
This symmetry is realized by introducing   a U(1) gauge field and a Newtonian prepotential, with which 
it can be shown that the  spin-0 graviton is eliminated \cite{HMT,WWc}. As a result, the  instability problem does not exist in this setup.
Another remarkable feature  of the setup is that it 
forces the coupling constant $\lambda$ to take exactly  its relativistic value $\lambda_{GR} = 1$.  Since both  the ghost and strong coupling problems
 are due precisely  to the deviation of $\lambda$ from   $1$,   this  implies that   these two problems   are also resolved.  
 
 However,  it  has been  argued  \cite{Silva}  that the introduction of the   Newtonian prepotential is so strong
that actions with $\lambda \not=1$ also have  the $U(1) \ltimes {\mbox{Diff}}(M, \; {\cal{F}})$ symmetry. 
Although the spin-0 graviton is still eliminated  for $\lambda \not=1$,  as shown explicitly 
by da Silva 
 for de Sitter and anti-de Sitter backgrounds \cite{Silva},  and Huang and Wang  for the Minkowski  \cite{HW}, the  ghost and strong coupling problems
arise again.   Indeed, it was shown  \cite{HW} that to avoid the ghost problem,   $\lambda$ 
must  satisfy  the constraints, 
$$
\lambda \ge 1\;\;\; {\mbox{or}} \;\;\; \lambda < 1/3.
$$ 
In addition, the coupling becomes strong for processes
with energy   higher  than $M_{pl} |\lambda -1|^{5/4}$ in the flat Friedmann-Robertson-Walker (FRW) background, and $M_{pl}|\lambda -1|^{3/2}$  in a 
static weak gravitational field.  It should be noted that in both cases to have non-vanishing   gravitational perturbations,  matter fields are necessarily 
present  \cite{HW}.

To solve the strong coupling problem \cite{CNPS}, two different approaches have been proposed. One is the BPS mechanism \cite{BPS}, in which a UV cutoff  
$M_{*}$ is introduced. By properly choosing the coupling constants involved in the theory, BPS showed that $M_{*}$ can be
lower than  $\Lambda_{SC}$, where $\Lambda_{SC}$ denotes the strong coupling energy scale of the theory. Then,
for processes with energies higher than $M_{*}$, high order derivative terms need to be taken into account.
The presence of these terms changes the scalings of the theory. In particular, 
all the irrelevant (nonrenormalizable) terms are turned  into either marginal (strictly renormalizable) or relevant
(superrenormalizable)  ones.
 As a result, the would-be strong coupling scale $\Lambda_{SC}$ disappeared,  due to the effects of high order derivative terms, and
  the theory becomes renormalizable \footnote {While this seems a very attractive 
mechanism, it turns out  \cite{WWb}  that it cannot be applied to the  Sotiriou-Visser-Weinfurtner (SVW) generalization \cite{SVW} (See also \cite{KKa}), because the instability of
the spin-0 graviton \cite{WWb}. However,  in the HMT setup, the Minkowski spacetime is stable, and the BPS mechanism now may become available.}.
The other approach is to provoke the  Vainshtein mechanism \cite{Vain}, as  showed recently in the spherical  static  \cite{Mukc} and
 cosmological \cite{WWb} spacetimes in the SVW setup \cite{SVW}.

\begin{figure}[tbp]
\centering
\includegraphics[width=8cm]{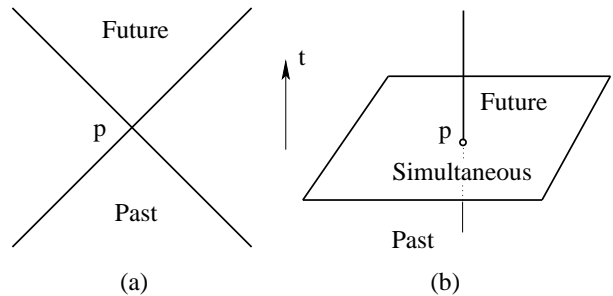}
\caption{ (a) The light cone of the event $p$ in special relativity. (b) The causal structure of the point $p$  in Newtonian theory. }
\label{fig0}
\end{figure}

In this paper, we leave the investigations of the strong coupling problem to another occasion, and   focus on another important issue: black holes in the HL theory.
In the HL theory, due to the breaking of the general covariance, the dispersion relations of particles  usually contain high order momentum terms
\cite{Mukc,Sotiriou,Padilla,Hreview,Visserb},
\bq
\lb{1.4}
\omega_{k}^{2} = m^{2} + k^{2}\left(1 + \sum^{z-1}_{n=1}{\lambda_{n}\left(\frac{k}{M_{n}}\right)^{2n}}\right),
\eq
for which the group velocity is given by \cite{CH}
\bq
\lb{1.4b}
v_{k} = \frac{k}{\omega}\left(1 + \sum^{z-1}_{n=1}{(n+1)\lambda_{n}\left(\frac{k}{M_{n}}\right)^{2n}}\right).
\eq
As an immediate result,  the speed of light becomes unbounded in the UV. This makes the causal
structure of the spacetimes quite different from that given in GR,  where the light cone of a given point $p$ plays a fundamental 
role in determining the causal relationship of $p$ to other events [cf. Fig. \ref{fig0}]. 
However, once the general covariance is broken, the causal 
structure will be dramatically changed. For example, in the Newtonian theory,  time is absolute and the speeds of signals are not limited. Then, the
causal structure of  a given point $p$ is uniquely determined by the time difference, $\Delta{t} \equiv t_{p} - t_{q}$, between the two events.  
 In particular, if $\Delta{t} > 0$, the event $q$ is to the past of $p$; if $\Delta{t} < 0$, it  is to the future; and if $\Delta{t} = 0$, the two events are
simultaneous.  

Another  consequence of the breaking of the general covariance is that  free particles now do not follow geodesics. This immediately makes
all the  definitions of black holes given in GR invalid \cite{HE73,Tip77,Hay94,Wang}.  To provide a proper definition of black holes, 
anisotropic conformal boundaries \cite{HMT2} and  kinematics of particles  \cite{KM} have been studied   within the HL framework. In this paper, we shall adopt the 
approach of Kiritsis and Kofinas (KK) \cite{KKb}, where a horizon is defined as the infinitely redshifted 2-dimensional (closed) surface of massless test
particles.  Clearly, such a definition reduces to that given in GR when the dispersion relation is relativistic [Where  $\lambda_{n} = 0$, as shown in 
Eq.(\ref{1.4}).]. 

 It should be noted that black holes in the HL theory with or without the projectability condition have been extensively studied, 
mainly using the definition borrowed directly from GR. In this paper, we shall show explicitly how these definitions are changed by considering some
 particular examples, found in the HMT set up with $\lambda = 1$. 
 
 Another  interesting approach is the equivalence  between   the HL theory (without the projectability
 condition) and the  Einstein-aether theory    in the IR  \cite{Jaco}, where the former is equivalent to the latter for the case where the aether vector field
 $u_{\mu}$ is hypersurface-orthogonal \footnote{In the spherically symmetric case, this is not a restriction as the aether field
 $u_{\mu}$ now is always hypersurface-orthogonal.}. From such studies one already sees the difficulties to define black holes, because of the fact that 
 different modes may have different 
 velocities even in the IR. In \cite{Jaco}, black holes are defined to possess both a metric horizon and a spin-0 mode horizon.   
 Since the equivalence holds only in the IR, it is still unclear how to extend such definitions  to high energy scales, where high order  curvature
  terms become important. 
 
  Specifically,  the paper is organized as follows: In Sec. II we briefly review the HMT setup (with $\lambda = 1$), while in
 Sec. III, we consider spherically symmetric black holes in the HL theory with  the KK approach \cite{KKb}. To keep our formulas  as  applicable as
 possible, in only this   section we consider spacetimes that may or may not satisfy the projectability condition.  We find that horizons are in general
 observer-dependent, and that with sufficient high energy, the radius of a horizon can be made arbitrarily small. This is consistent with  the fact 
 that the speed of light now becomes unbounded in the UV.  In Sec. IV, we study all the vacuum diagonal ($N^{i} = 0$) 
 solutions obtained in  the HMT setup \cite{HMT,AP,GSW}, paying particular attention to their global structures.  
 Using the definition of horizons, we study their existence in various cases.  It is remarkable that in some cases the structure of the Einstein-Rosen bridge 
exists, where a throat with finite non-zero radius connects two asymptotically flat regions.   Due to the restricted diffeomorphisms (\ref{1.3}), Penrose diagrams are not
 allowed. However, for the sake of comparison, we present the corresponding Penrose diagrams obtained by assuming that the general general transformations are still allowed. 
 In Sec. V, we study  the nondiagonal ($N^{i}\not= 0$) vacuum solutions obtained in \cite{AP,GSW}, while in   Sec. VI, our main conclusions are presented. 
 There are also two Appendices, A and B. In Appendix A, the 3-tensor $F_{ij}$ for the spherical  spacetimes are given, while in Appendix B, we study
  slowly-rotating solutions in the HMT setup, and obtain all the solutions, which includes    the Kerr solution  given in GR.

\section{Nonrelativistic general covariant HL theory}

\renewcommand{\theequation}{2.\arabic{equation}} \setcounter{equation}{0}

The  nonrelativistic general covariant HL theory  is described by the action \cite{HMT,WWc},
 \bqn \lb{2.4}
S &=& \zeta^2\int dt d^{3}x N \sqrt{g} \Big({\cal{L}}_{K} -
{\cal{L}}_{{V}} +  {\cal{L}}_{{\varphi}} +  {\cal{L}}_{{A}} \nb\\
& & ~~~~~~~~~~~~~~~~~~~~~~ \left. + {\zeta^{-2}} {\cal{L}}_{M} \right),
 \eqn
where $g={\rm det}\,g_{ij}$, and
 \bqn \lb{2.5}
{\cal{L}}_{K} &=& K_{ij}K^{ij} -    K^{2},\nb\\
{\cal{L}}_{\varphi} &=&\varphi {\cal{G}}^{ij} \Big(2K_{ij} + \nabla_{i}\nabla_{j}\varphi\Big),\nb\\
{\cal{L}}_{A} &=&\frac{A}{N}\Big(2\Lambda_{g} - R\Big). 
 \eqn
Here  $\Lambda_{g}$ is a    coupling constant and 
 \bqn \lb{2.6}
K_{ij} &=& \frac{1}{2N}\left(- \dot{g}_{ij} + \nabla_{i}N_{j} +
\nabla_{j}N_{i}\right),\nb\\
{\cal{G}}_{ij} &=& R_{ij} - \frac{1}{2}g_{ij}R + \Lambda_{g} g_{ij},
 \eqn
where the Ricci   terms all refer to the three-metric $g_{ij}$.
${\cal{L}}_{M}$ is the matter Lagrangian density,   and 
${\cal{L}}_{{V}}$ is a  Diff($\Sigma$)-invariant local scalar functional. With the assumptions that  the highest order 
derivatives are six, and   the parity is conserved,  ${\cal{L}}_{{V}}$ takes the general  form \cite{SVW}, 
 \bqn \lb{2.5a} 
{\cal{L}}_{{V}} &=& \zeta^{2}g_{0}  + g_{1} R + \frac{1}{\zeta^{2}}
\left(g_{2}R^{2} +  g_{3}  R_{ij}R^{ij}\right)\nb\\
& & + \frac{1}{\zeta^{4}} \left(g_{4}R^{3} +  g_{5}  R\;
R_{ij}R^{ij}
+   g_{6}  R^{i}_{j} R^{j}_{k} R^{k}_{i} \right)\nb\\
& & + \frac{1}{\zeta^{4}} \left[g_{7}R\nabla^{2}R +  g_{8}
\left(\nabla_{i}R_{jk}\right)
\left(\nabla^{i}R^{jk}\right)\right],  ~~~~
 \eqn 
 where the coupling  constants $ g_{s}\, (s=0, 1, 2,\dots 8)$  are all dimensionless. The relativistic limit in the IR
 requires $g_{1} = -1$ and $\zeta^2 = 1/(16\pi G)$. 

Then, it can be shown that the  Hamiltonian and momentum constraints are given respectively by,
 \bqn 
 \lb{eq1}
& & \int{ d^{3}x\sqrt{g}\Big({\cal{L}}_{K} + {\cal{L}}_{{V}} - \varphi {\cal{G}}^{ij}\nabla_{i}\nabla_{j}\varphi\Big)}\nb\\
& & ~~~~~~~~~~~~~~~~
= 8\pi G \int d^{3}x {\sqrt{g}\, J^{t}},\\
\lb{eq2}
& & \nabla^{j}\Big(\pi_{ij} - \varphi  {\cal{G}}_{ij}\Big) = 8\pi G J_{i},
 \eqn
where
 \bqn 
  \lb{eq2b}
  J^{t} &\equiv& 2 \frac{\delta\left(N{\cal{L}}_{M}\right)}{\delta N},\nb\\
   \pi_{ij} &\equiv& 
   - K_{ij} +   K g_{ij},\nb\\
 J_{i} &\equiv& - N\frac{\delta{\cal{L}}_{M}}{\delta N^{i}}.
 \eqn

Variation of the action (\ref{2.4}) with respect to   $\varphi$ and $A$ yield, 
\bqn
\lb{eq4a}
& & {\cal{G}}^{ij} \Big(K_{ij} + \nabla_{i}\nabla_{j}\varphi\Big)  = 8\pi G J_{\varphi}, \\
\lb{eq4b}
& & R - 2\Lambda_{g} =   8\pi G J_{A},
\eqn
where
\bq
\lb{eq5}
J_{\varphi} \equiv - \frac{\delta{\cal{L}}_{M}}{\delta\varphi},\;\;\;
J_{A} \equiv 2 \frac{\delta\left(N{\cal{L}}_{M}\right)}{\delta{A}}.
\eq
On the other hand,  the dynamical equations now read,
 \bqn \lb{eq3}
&&
\frac{1}{N\sqrt{g}}\Bigg[\sqrt{g}\Big(\pi^{ij} - \varphi {\cal{G}}^{ij}\Big)\Bigg]_{,t}  
\nb\\
& &~~~ = -2\left(K^{2}\right)^{ij}+2 K K^{ij} \nb\\
& &  ~~~~~ + \frac{1}{N}\nabla_{k}\left[N^k \pi^{ij}-2\pi^{k(i}N^{j)}\right]\nb\\ 
& & ~~~~~ +  \frac{1}{2} \Big({\cal{L}}_{K} + {\cal{L}}_{\varphi} + {\cal{L}}_{A}\Big) g^{ij} \nb\\
& &  ~~~~~    + F^{ij} + F_{\varphi}^{ij} +  F_{A}^{ij} + 8\pi G \tau^{ij},
 \eqn
where $\left(K^{2}\right)^{ij} \equiv K^{il}K_{l}^{j},\; f_{(ij)}
\equiv \left(f_{ij} + f_{ji}\right)/2$, and
 \bqn
\lb{eq3a} 
F^{ij} &\equiv&
\frac{1}{\sqrt{g}}\frac{\delta\left(-\sqrt{g}
{\cal{L}}_{V}\right)}{\delta{g}_{ij}}
 = \sum^{8}_{s=0}{g_{s} \zeta^{n_{s}}
 \left(F_{s}\right)^{ij} },\nb\\
F_{\varphi}^{ij} &=&  \sum^{3}_{n=1}{F_{(\varphi, n)}^{ij}},\nb\\
F_{\varphi}^{i} &=&  \Big(K + \nabla^{2}\varphi\Big)\nabla^{i}\varphi + \frac{N^{i}}{N} \nabla^{2}\varphi, \nb\\
F_{A}^{ij} &=& \frac{1}{N}\left[AR^{ij} - \Big(\nabla^{i}\nabla^{j} - g^{ij}\nabla^{2}\Big)A\right], 
 \eqn
 where $n_{s} =(2, 0, -2, -2, -4, -4, -4, -4,-4)$,  and the geometric 3-tensors $ \left(F_{s}\right)_{ij}$ and $F_{(\varphi, n)}^{ij}$ are given
in \cite{WWc}.  The
stress 3-tensor $\tau^{ij}$ is defined as
 \bq \label{tau}
\tau^{ij} = {2\over \sqrt{g}}{\delta \left(\sqrt{g}
 {\cal{L}}_{M}\right)\over \delta{g}_{ij}}.
 \eq

The matter quantities $(J^{t}, \; J^{i},\; J_{\varphi},\; J_{A},\; \tau^{ij})$ satisfy the
conservation laws,
 \bqn \lb{eq5a} & &
 \int d^{3}x \sqrt{g} { \left[ \dot{g}_{kl}\tau^{kl} -
 \frac{1}{\sqrt{g}}\left(\sqrt{g}J^{t}\right)_{, t}  
 +   \frac{2N_{k}}  {N\sqrt{g}}\left(\sqrt{g}J^{k}\right)_{,t}
  \right.  }   \nb\\
 & &  ~~~~~~~~~~~~~~ \left.   - 2\dot{\varphi}J_{\varphi} -  \frac{A} {N\sqrt{g}}\left(\sqrt{g}J_{A}\right)_{,t}
 \right] = 0,\\
\lb{eq5b} & & \nabla^{k}\tau_{ik} -
\frac{1}{N\sqrt{g}}\left(\sqrt{g}J_{i}\right)_{,t}  - \frac{J^{k}}{N}\left(\nabla_{k}N_{i}
- \nabla_{i}N_{k}\right)   \nb\\
& & \;\;\;\;\;\;\;\;\;\;\;- \frac{N_{i}}{N}\nabla_{k}J^{k} + J_{\varphi} \nabla_{i}\varphi - \frac{J_{A}}{2N} \nabla_{i}A
 = 0.
\eqn

 \section{Black Holes in HL Theory }

\renewcommand{\theequation}{3.\arabic{equation}} \setcounter{equation}{0}

KK considered a scalar field with a given dispersion relation $F(\zeta)$   \cite{KKb}. In the geometrical optical  approximations,  
$\zeta$ is given by $\zeta = g_{ij}k^{i}k^{j}$, where $k_{i}$ denotes the 3-momentum of the corresponding spin-0 particle. With this  approximation, 
the trajectory of a test particle is given by
\bqn
\lb{eqq1}
S_{p} &\equiv& \int_{0}^{1}{{\cal{L}}_{p} d\tau} \nb\\
&=& \frac{1}{2} \int_{0}^{1}{ d\tau\Bigg\{\frac{c^{2}N^{2}}{e} \dot{t}^{2} + e \Big[F(\zeta) - 2 \zeta F'(\zeta)\Big]\Bigg\}}, ~~~~
\eqn
where $e$ is a one-dimensional einbein, and   $\zeta$ is now considered as a functional of $t, x^{i}, \dot{t}, \dot{x}^{i}$ and $e$,  given by the relation,
\bq
\lb{eqq2}
\zeta\;  [F'(\zeta)]^{2} = \frac{1}{e^{2}} g_{ij}\big(\dot{x}^{i} + N^{i} \dot{t}\big) \big(\dot{x}^{j} + N^{j}  \dot{t}\big),
\eq
with $\dot{t} \equiv dt/d\tau$, etc.  For detail, we refer readers to \cite{KKb}.

It should be noted that KK obtained the above action starting from a scalar field. So, strictly speaking, it is valid only for spin-0 test particles. However, what is really 
important in their derivations is the dispersion relationship $F(\zeta)$. As shown in \cite{Wangb}, a spin-2 particle  has a similar dispersion relation. It is
expected that a spin-1 test particle, such as photons,  should have a similar dispersion relation too \cite{CH,KKb}. Therefore, in the rest of this paper and without proof,
we simply consider the action (\ref{eqq1})  to describe all massless test particles. 

Spherically symmetric static spacetimes in the framework of the HMT setup were studied systematically in \cite{AP,GSW},
and  the metric for  static spherically symmetric spacetimes that preserve the form of Eq. (\ref{1.2}) 
with the projectability condition can be cast in the form \cite{GPW} \footnote{Note the slight difference between the $g_{tr}$ term defined here and the one
defined in \cite{GPW,GSW}.},   
\bq
\lb{3.1b}
ds^{2} = - c^{2}dt^{2} + e^{2\nu} \left(dr + e^{\mu - \nu} cdt\right)^{2}  + r^{2}d^2\Omega,  
\eq
  where  $d^2\Omega = d\theta^{2}  + \sin^{2}\theta d\phi^{2}$,
and 
\bq
\lb{3.1}
 \mu = \mu(r),\;\;\; \nu = \nu(r),\;\;\; N^{i} = \left\{ce^{\mu - \nu}, 0, 0\right\}.
 \eq
The corresponding timelike Killing vector is  $\xi = \partial_{t}$, and the diagonal case $N^{r} = 0$ corresponds to $\mu = -\infty$.

However, to study black hole solutions in a more general case, in this (and only in this) section, we also consider the case without projectability condition, and write
  the metric as,
\bq
\lb{eqq3}
ds^{2} = - N^{2}c^{2}dt^{2} + \frac{1}{f} \left(dr + N^{r} c dt\right)^{2}  + r^{2}d^2\Omega. 
\eq
where $N,\; f$ and $N^{r}$ are all functions of $r$.  
Without loss of generality, in the rest of the paper we shall set $c = 1$, 
which is equivalent to the coordinate transformations
$x_{0} = c t,\; \bar{N}^{r} = N^{r}/c$. Taking
\bq
\lb{F-function}
F(\zeta) = \zeta^{n}, \; (n = 1, 2, ...),
\eq
Eq.(\ref{eqq2}) yields,
\bq
\lb{eqq4}
\zeta = \left(\frac{\dot{r} + N^{r}\dot{t}}{n e \sqrt{f}}\right)^{{2}/{(2n - 1)}} \equiv \left(\frac{{\cal{D}}}{e^{2}}\right)^{{1}/{(2n - 1)}}.
\eq
Inserting this into Eq.(\ref{eqq1}), we find that, for  radially moving particles, ${\cal{L}}_{p}$ is given by
\bq
\lb{eqq5}
{\cal{L}}_{p} = \frac{N^{2}}{2e} \dot{t}^{2} + \frac{1}{2} \big(1 - 2n\big)e^{1/(1-2n)} {\cal{D}}^{{n}/{(2n - 1)}}.
\eq
Then, from the equation $\delta{\cal{L}}_{p}/\delta{e} = 0$ we obtain
\bq
\lb{eqq6}
N^{2}\dot{t}^{2} - e^{2(n-1)/(2n-1)} {\cal{D}}^{{n}/{(2n - 1)}} = 0.
\eq
On the other hand, since $\delta{\cal{L}}_{p}/\delta{t} = 0$, the Euler-Lagrange equation,
$$
\frac{\delta{\cal{L}}_{p}}{\delta{t}} - \frac{1}{d\tau}\left(\frac{\delta{\cal{L}}_{p}}{\delta{\dot{t}}}\right) = 0,
$$ 
yields
\bq
\lb{eqq8}
N^{2}\dot{t} - e^{2(n-1)/(2n-1)}\frac{N^{r}}{\sqrt{f}} {\cal{D}}^{{1}/{[2(2n - 1)]}} = e E,
\eq
where $E$ is an integration constant, representing the total energy of the test particle. 

To solve Eqs.(\ref{eqq6}) and (\ref{eqq8}),   we first consider the case $n = 1$, which  corresponds to the relativistic dispersion relation. 
From such considerations, we shall see how to generalize the definition of black holes given in GR to the HL theory where $n$ is generically 
different from 1, as required by the renornalizability condition in  the UV.

\subsection{$ n = 1$}

In this  case,  Eqs.(\ref{eqq6}) and (\ref{eqq8}) reduce, respectively, to,
\bqn
\lb{eqq9a}
N^{2}\dot{t}^{2} - {\cal{D}} &=& 0,\\
\lb{eqq9b}
N^{2}\dot{t}  -  {N^{r}}\sqrt{\frac{\cal{D}}{f}}  &=& e E.
\eqn
Eq.(\ref{eqq9a}) simply tells us that now the particle moves along   null  geodesics. The above equations  can be
easily solved according to whether $N^{r}$ vanishes or not. 

 \subsubsection{$N^{r} = 0$}
 
 When $N^{r} = 0$, from Eq.(\ref{eqq9a}) we find
 \bq
 \lb{eqq10}
 dt = \pm \frac{dr}{N \sqrt{f}},
 \eq
where ``+" (``$-$") corresponds to out-going (in-going) light rays. If $f$ has an a-th order zero and $N^{2}$ a b-th order zero at a surface, say, $r = r_{g}$, that is,
\bq
\lb{fN}
f = f_{0}(r)(r - r_{g})^{a},\;\;\;
N = N_{0}(r) (r - r_{g})^{b/2}, 
\eq
where  $N_{0}(r_{g}) \not= 0$ and $f_{0}(r_{g}) \not= 0$, then from the above
equations we find that in the neighborhood of $r = r_{g}$, 
\bq
\lb{eqq11}
t \simeq t_{0} \pm \frac{1}{N_{0}\sqrt{f_{0}}}\cases{\frac{2}{2 - (a+b)}(r-r_{g})^{1 - (a+b)/2}, & $a+b \not= 2$,\cr
\ln\left|r-r_{g}\right|, & $a +b  =2$.\cr}
\eq
Therefore, when
\bq
\lb{eqq12}
a + b \ge 2, \; (n = 1),
\eq
$t$  becomes unbounded, 
as $r \rightarrow r_{g}$, 
at which 
the light rays are infinitely redshifted. 
This indicates that an event  horizon might exist at $r = r_{g}$, provided that the spacetime has no curvature singularity there. 
A simple example is the Schwarzschild solution, $N^{2} = f = (r - r_{g})/r$, which is also a solution of the HL theory without the projectability condition, but with
the  detailed balance condition  softly broken \cite{KK},
and for which we have $a = b = 1$. Clearly, it satisfies  the above condition with the equality, so $r = r_{g}$ indeed defines a horizon. 

 \subsubsection{$N^{r} \not = 0$}
 
 When $N^{r} \not = 0$, Eq.(\ref{eqq9a}) yields
 \bq
 \lb{eqq13}
 t = t_{0} +  \int{\frac{\epsilon dr}{N\sqrt{f} - \epsilon N^{r}}},
 \eq
where $\epsilon = +1\; (\epsilon = - 1)$ corresponds to out-going (in-going) light rays. If 
\bq
\lb{eqq13aaa}
H(r) \equiv N\sqrt{f} - \epsilon N^{r},
\eq
 has $\delta$-th order zero at $r_{g}$, 
\bq
\lb{H-functin}
H(r) = H_{0}(r)(r - r_{g})^{\delta},
\eq
with $H_{0}(r_{g}) \not= 0$, 
we find that in the neighborhood $r = r_{g}$ Eq.(\ref{eqq13}) yields
\bq
\lb{eqq14}
t = t_{0} +\frac{\epsilon}{H_{0}(r_{g})}\cases{\frac{1}{1-\delta}(r-r_{g})^{1-\delta}, & $ \delta \not= 1$,\cr
\ln(r-r_{g}), & $\delta = 1$.\cr}
\eq
Clearly, when 
\bq
\lb{eqq15}
\delta \ge 1, (n = 1), 
\eq
$|t|$ becomes unbounded as $r \rightarrow r_{g}$, and  an event horizon might exist. 

The Schwarzschild solution in the Painlev\'e-Gullstrand coordinates  \cite{GP} is given by 
\bq
\lb{Sch}
N_{Sch}^{2} = f_{Sch}  = 1,\;\;\;
N_{Sch}^{r} = \epsilon_{1} \sqrt{\frac{r_{g}}{r}},
\eq
 where
$\epsilon_{1} = \pm 1$.  As shown in \cite{AP,GSW}, this is also a vacuum solution of the HL theory in the HMT setup \cite{HMT}. Then, we find that $H(r) = 1 - \epsilon_{1}\epsilon  \sqrt{r_{g}/r}$. 
Thus, for the solution with  $ \epsilon_{1} = +1$, the time
of  the out-going null rays, measured by asymptotically flat observers,   
becomes unbounded at $r_{g}$, and for  the solution with  $ \epsilon_{1} = -1$, the time
of  the in-going null rays   becomes unbounded. Therefore,  an event horizon is indicated to exist at $r = r_{g}$ in both cases. 

 In review of the above, KK generalized the notion of black holes defined in GR to the case of a  non-standard dispersion relation \cite{KKb}. 
 In summary, {\em a horizon is defined as a surface on which light rays are infinitely redshifted}. It should be noted that this redshift should be understood as measured by asymptotically 
 flat observers  at $N(r \gg r_{g}) \simeq 1$ and $N^{r}(r\gg r_{g}) \simeq 0$, with $r$ being the geometric radius, $r = \sqrt{A/4\pi}$, of the 2-sphere:  $t, r = $ Constants,  where $A$
 denotes the area of the 2-sphere.

 
 \subsection{$n \ge 2$}
 
 In this case, eliminating $e$ from Eqs.(\ref{eqq6}) and (\ref{eqq8}) we find that
 \bq
 \lb{eqq16}
 X^{n} - p(r) X - q(r, E) = 0,
 \eq
 where
 \bqn
 \lb{eqq17}
 X &\equiv& \left(\frac{\sqrt{\cal{D}}}{\dot{t}}\right)^{1/(n-1)} = \left(\frac{\left|r' + N^{r}\right|}{n \sqrt{f}}\right)^{1/(n-1)} ,\nb\\
 p(r) &\equiv& \frac{N^{r}}{\sqrt{f}},\;\;\;
 q(r, E) \equiv E N^{1/(n-1)}, 
 \eqn
 with $r' \equiv \dot{r}/\dot{t} = dr/dt$. To solve the above equation, again it is found convenient to consider the cases $N^{r} = 0$ and $N^{r} \not=0$
 separately. 
 
 \subsubsection{$N^{r} = 0$}
 
 When $N^{r} = 0$, Eq.(\ref{eqq16}) has the solution,
 \bq
 \lb{eqq18}
 t = t_{0} + \epsilon \int{\frac{dr}{n E^{(n-1)/n}\sqrt{f}N^{1/n}}},
 \eq
 where $\epsilon = +1$ corresponds to outgoing rays, and $\epsilon = -1$   to  ingoing rays.
 Thus, if  $f$ has an a-th order zero and $N^{2}$ a b-th order zero at $r = r_{g}$,  as given by Eq.(\ref{fN}),
 we have $\sqrt{f}N^{1/n}\sim (r-r_{g})^{(a + b/n)/2}$. Then, from the above, we find that
  the time $t$, measured by asymptotically flat observers, becomes infinitely large
 at $r = r_{g}$, provided that  \cite{KKb}
 \bq
 \lb{eqq19}
 a + \frac{b}{n} \ge 2.
 \eq
  For the solutions with the projectability condition ($N = 1,\; b = 0$), this is possible only when $a \ge 2$. 
  
  Considering  again 
  the Schwarzschild solution, $N^{2} = f = (r - r_{g})/r$, one finds  that this does not satisfy the condition (\ref{eqq19}) 
  with $n \ge 2$. Therefore, the  Schwarzschild black hole in GR is no longer a black hole in the HL theory, because of the non-relativistic dispersion relations
  (\ref{1.4}).
  This is expected, since even in GR when quantum effects are taken into account, such as the Hawking radiation, classical black holes are no longer
  black. 

    \subsubsection{$N^{r} \not= 0$}
    
   In this case,  let us consider an ingoing ray ${r}'< 0$. Suppose there is a horizon located at $r = r_{H}$. Then $r'(r) \simeq 0$ as we approach the horizon. Thus, if $N^r > 0$ and bounded away from zero,   $(r' + N^r)$ will also be positive, when the ray is sufficiently near the horizon. Conversely, if $N^r < 0$ and bounded away from zero, then $(r' + N^r)$ will  be negative sufficiently near the horizon.
Defining $H$ by $H(r, E) \equiv r'$, we find that for an ingoing ray near the horizon we have,
\bqn
\lb{4.18}
& t = t_0 + \int \frac{dr}{H(r,E)},
	\\ 
\lb{4.19}
 & H(r,E) = \epsilon n \sqrt{f}X^{n-1} - N^r,
\eqn
where
\bq
\epsilon = \cases{1, & $N^r > 0$, \cr
-1, & $ N^r < 0$. \cr
}
\eq
Dividing (\ref{eqq16}) by $X$ and solving for $X^{n-1}$, we obtain
$$
X^{n-1} = \frac{N^r}{\sqrt{f}} + \frac{EN^{\frac{1}{n-1}}}{X}.
$$
Substituting  this   into (\ref{4.19}), we find,
\bq
\lb{*}
  H = (\epsilon n - 1) N^r + \epsilon n \sqrt{f}\frac{EN^{\frac{1}{n-1}}}{X}.
\eq
It follows that if $H$ has a zero at $r = r_{H}$, then
\bq
\lb{**}
  X|_{r = r_{H}} = - \frac{\epsilon n \sqrt{f} EN^{\frac{1}{n-1}}}{(\epsilon n - 1) N^r}.
\eq
The expression on the rhs is positive (negative) for $\epsilon = -1$ ($\epsilon = 1$). Thus, $H$ can have a zero only if $\epsilon = -1$.
Thus, we will henceforth consider only this case.
Differentiation of (\ref{*}) with respect to $r$ yields
\bqn
\lb{***}
H'(r) &=& - (n + 1)N^{r'} - n \sqrt{f}\frac{EN^{\frac{1}{n-1} - 1}}{(n-1)X}N' \nb\\
& & 
- n \frac{EN^{\frac{1}{n-1}}}{2 \sqrt{f}X}f'  +  n \sqrt{f}\frac{EN^{\frac{1}{n-1}}}{X^2}X'. ~~~~
\eqn
On the other hand, differentiation of (\ref{eqq16}) with respect to $r$ yields
\bqn
X'(r)  &=&  \frac{1}{nX^{n-1} - \frac{N^r}{\sqrt{f}}} \left[\left(\frac{1}{\sqrt{f}} \frac{dN^r}{dr} - \frac{N^r}{2f^{3/2}} \frac{df}{dr}\right)X\right. \nb\\
& & \left. + \frac{EN^{\frac{1}{n-1} -1}}{n-1}\frac{dN}{dr}\right].
\eqn
Substituting the above into Eq.(\ref{***}), we find that
\bqn
& &H'(r) \bigg|_{r = r_{H}} =
-\frac{n+1}{2} \Biggl(\frac{H_{1}}{H_{2}} -\frac{N^r f'}{f}
   	\nb\\
&&   ~~~~~~~~~~~~~~~~~~
   +\frac{2 N^rN'}{N-n N}+2 N^{r'}\Biggr),
   \eqn
where
\bqn
& &  H_{1}\equiv 2 E (n+1) f N^r N' \nb\\
 & & ~~~~~~~ + E (n-1) n N \left(N^r f' - 2 f N^{r'}\right),\nb\\
& &  H_{2} \equiv (n-1) n f N \Bigg[E + (n+1) N^{\frac{1}{1-n}}\nb\\
   & & ~~~~~~~  \times \left(\frac{E n \sqrt{f}
   N^{\frac{1}{n-1}}}{(-n-1) N^r}\right)^n\Bigg].
\eqn

If $H$ has a zero of order $\delta > 0$ at $r_{H}$, we can write it in the form, 
\bq
\lb{H-functionb}
H(r) = H_0(r_{H})(r - r_{H})^\delta + \cdots,  
\eq
as $r \to r_{H}$, where $H_0(r_{H}) \neq 0$.  
Therefore,
\bq
\lb{CT}
H'(r) \bigg|_{r = r_{H}} = \cases{ 0, & $ \delta > 1$, \cr
H_0(r_{H}), & $ \delta  = 1$, \cr
\pm \infty, & $0 <\delta < 1$.}
\eq
Now $t \to \infty$ as $r \rightarrow  r_{H}^{+}$ if and only if 
\bq
\lb{delta}
\delta \geq 1, 
\eq
which happens if and only if ${dH}/{dr} \big|_{r = r_{H}}$ is  finite. 
This gives an explicit condition on $f, N, N^r, E, n$ for the blow-up of $t$ at $r_{H}$.

It should be noted that $r_{H}$ usually depends on the energy $E$ of the test particles, as can be seen 
from the above and specific examples considered below. 

%

{\bf Case $n=2$:} In this case, we have
\bq
H'(r)\bigg|_{r = r_{H}} =
\frac{H_{3}}{2 N \left[4 E f N+3 (N^r)^2\right]}, 
\eq
where
\bqn
& &   H_{3} \equiv  3 \left[4 E N^2 \left(N^r f'-2 f
   N^{r'}\right)+8 E f N N^r N'\right. \nb\\
   & &  ~~~~~~~~~~~ -3 (N^r)^3 N'\Big], (n = 2).
 \eqn
Again, for the Schwarzschild solution (\ref{Sch}),  
we have
\bqn
X(r) &=& \frac{1}{2} \left( - \sqrt{\frac{r_g}{r}} + \frac{\sqrt{4 E r+ r_g}}{\sqrt{r}}\right), \nb\\
H(r) &=& \frac{-3 \sqrt{r_g (4 E r+r_g)}+4 E r+3
   r_g}{\sqrt{r r_g}-\sqrt{r (4 E r+r_g)}}, 
  \eqn
 so that $H(r) = 0$ has the solution,
 \bq
 \lb{root} 
  r_{H} = \frac{3 r_g}{4E}, \; (n = 2),
  \eq
  at which we have
 \bq
  H'(r_{H}) = -\frac{2 E^{3/2}}{\sqrt{3} r_g}.
 \eq
 Then, according to Eq.(\ref{CT}), we have $\delta = 1$,  i.e.,  $t$ diverges logarithmically as $r \rightarrow  r_{H}^{+}$. Therefore, in this case
there does exist a horizon. But, the location of it  depends on the energy $E$ of the test particle, and approaches zero   when $E \gg r_{g}$.  
This is understandable, as  the speed of light is unbounded in the UV, and in principle the  singularity located at $r = 0$ 
can be seen by asymptotically flat observers, as long as the  light rays sent by the observers have sufficiently high energies.  

{\bf Case $n=3$:} In this case, we have
\bq
\lb{casen3}
X^{3} - p(r) X - q(r, E) = 0.
\eq
Assuming that $H(r) = 0$ has a real and positive root $r_{H}$, we find that
\bq
\lb{CT3}
H'(r) \bigg|_{r = r_{H}} = 
\frac{H_{4}}{3 N \left(27 E^2 f^{3/2} N-16 (N^r)^3\right)},
\eq
where
\bqn
& & H_{4} \equiv 162 E^2 \sqrt{f} N^2 N^r f'  +32 (N^r)^4 N'\nb\\
& & ~~~~~~~~  -162 E^2 f^{3/2} N \left(2 N N^{r'}-N^r N'\right). 
\eqn

For the Schwarzschild solution (\ref{Sch}),  
we have $p(r) = -\sqrt{r_g/r}$ and $q(r, E) = E$. Then,
we find that 
\bqn
\lb{casen3a}
& & X^{3} +\sqrt{\frac{r_g}{r}} X -E = 0,\\
\lb{casen3b}
&& H(r) = 4\sqrt{\frac{r_g}{r}} - \frac{3E}{X},
\eqn
from which  we find that $H (r) =0$ has a solution,  
\bq
\lb{rootb}
r_{H} = r_{g}\left(\frac{16}{27E^{2}}\right)^{2/3}, \; ( n = 3),
\eq
which  also depends on $E$, and approaches zero as $E \rightarrow \infty$. 
Substituting $r_{H}$  into Eq.(\ref{CT3}),  we find
$H'(r_{H}) = - 27E^{2}/(16r_{g})$. That is, the hypersurface $r = r_{H}$ is also an observer-dependent  horizon in the case $n = 3$, and the radius of the horizon is
inversely proportional to the energy of the test particle. For $E \gg r_{g}$, we have $r_{H} \simeq 0$.

Another (simpler)  consideration  for the existence of the horizon is given as follows: First, from Eq.(\ref{eqq17}) we find that
\bqn
\lb{JX1}
X &=& \left(\frac{|r' + N^r|}{n \sqrt{f}}\right)^{1/(n -
       1)} \simeq  \left(\frac{\epsilon N^r}{n\sqrt{f}}\right)^{1/(n - 1)}\nb\\
       & & ~~~~~~~~~~~~~~~~~~~
            \times \left(1 + \frac{H}{(n - 1) N^r}\right),  
\eqn
for $r \simeq r_{H}$. Inserting it into Eq.(\ref{*}), we have, to leading order,
\bqn
\lb{JX2}
&&  \left(1 + \frac{E N^{1/(n - 1)}}{(n - 1)
       \left(\frac{\epsilon N^r}{n \sqrt{f}}\right)^{n/(n -
       1)}}\right) H (r, E) = (\epsilon n - 1) N^r\nb\\
       & & ~~ ~~~~\times \left(1 + \frac{E N^{1/(n - 1)}}{(\epsilon n - 1)
       \left(\frac{\epsilon N^r}{n \sqrt{f}}\right)^{n/(n -
       1)}}\right).
\eqn
Then, we obtain
\bq
\lb{JX3}
\left.\frac{E N^{1/(n - 1)}}{( n + 1)
       \left(\frac{- N^r}{n \sqrt{f}}\right)^{n/(n -
       1)}}\right|_{r = r_H} = 1. 
\eq
 Given this, we can further simplify Eq.(\ref{JX2}) to, 
 \bqn
 \lb{JX4}
&&   \frac{2 n}{n - 1} H (r, E) = - (n + 1) N^r (r_H)\nb\\
& & ~~~~~\times
\left(1 - \frac{E N^{1/(n - 1)} (r)}{( n + 1)
       \left(\frac{- N^r (r)}{n \sqrt{f (r)}}\right)^{n/(n -
       1)}}\right). ~~~~~~~~~~
 \eqn
 Then, using Eq.(\ref{H-functionb}), we have the following constraint for $N, N^r, f$ to
       satisfy so that a horizon can indeed exist, 
\bqn
\lb{CDb}       
&&  \frac{E N^{1/(n - 1)} (r) }{( n + 1)
       \left(\frac{- N^r (r)}{n \sqrt{f (r)}}\right)^{n/(n -
       1)}} \nb\\
       & & ~~~ = 1 + \frac{2 n H_0 (r_H)}{(n^2 - 1) N^r (r_H)} (r -
       r_H)^\delta + \cdots. ~~~~~~~~
\eqn
This equation can be first used to determine $r_H$   and then   $\delta$,  once $N, \; N^r$ and $ f$ are given.  
To illustrate how to use it,  let us consider  the Schwarzschild metric (\ref{Sch}). For $n = 2$, $r_H$
 can  be obtained simply from the above, and is given   exactly by Eq.(\ref{root}), 
for which we have
\bqn
\lb{JX6}
&& \frac{E N^{1/(n - 1)} (r) }{( n + 1)
       \left(\frac{- N^r (r)}{n \sqrt{f (r)}}\right)^{n/(n -
       1)}} 
       \simeq 1 + \frac{r - r_H}{r_H}, ~~~~~
\eqn
that is, $\delta = 1$.

For $n = 3$, from Eq.(\ref{JX3}) we find that $r_{H}$ is given by Eq.(\ref{rootb}), and 
\bqn 
&& \frac{E N^{1/(n - 1)} (r) }{( n + 1)
       \left(\frac{- N^r (r)}{n \sqrt{f (r)}}\right)^{n/(n -
       1)}}  = \frac{3^{3/2} E r^{3/4}}{4 r_g^{3/4}}  \nb\\
       & &~~~~~~~~~~ \simeq 1 +
       \frac{3}{4 r_H} (r - r_H) + \cdots.
\eqn
Therefore, in this case we have  $\delta = 1$ too.

 It should be noted that in the above analysis, we assumed that $F(\zeta) = \zeta^{n}$. In  more realistic models, the dispersion relation is a polynomial of
 $\zeta$, as shown by Eq.(\ref{1.4}), or more specifically, 
 \bq
 \lb{Poly}
 F(\zeta) = \zeta + \frac{\zeta^{2}}{M^{2}_{A}}  +   \frac{\zeta^{4}}{M^{4}_{B}} + ...,
 \eq
 where $M_{A}$ and $M_{B}$ are the energy scales, which can be significantly different from the Planck one \cite{BPS}. Therefore, for observers in low energy scales,
 where $\zeta \ll M_{A}, M_{B}$, the first term dominates, and some solutions, including the Schwarzschild solution, look like black holes, as shown in the case $n = 1$.
But,  for observers with high energies, those solutions may not be black holes any longer. Even if they are,  their horizons in general are observer-dependent,  as shown in the 
cases $n = 2$ and $n = 3$ explicitly for the Schwarzschild solution. To illustrate the main properties of the dispersion relation (\ref{Poly}),  we shall consider 
the case where only the first two terms are  important. 

\subsection{ Trajectories of Test Particles with the Dispersion Relation $F(\zeta) = \zeta + {\zeta^{2}}/{M^{2}_{A}}$}

For the sake of simplicity, we  restrict ourselves to the  case $N^{r} = 0$. 
Substituting 
\bq
 \lb{Polya}
 F(\zeta) = \zeta + \frac{\zeta^{2}}{M^{2}_{A}},
 \eq
  into  Eq.(\ref{eqq2}), we find 
\bq
\label{zetaeq}
  \zeta \left(1 + \frac{2\zeta}{M_A^2}\right)^2 = \frac{\dot{r}^2}{e^2f}.
\eq
Solving this equation directly for $\zeta$ yields a very complicated expression, and it is not clear how to proceed along this direction. 
Instead, we note that our goal is to find the analog of equation (\ref{eqq6}), i.e. of the equation $\delta \mathcal{L}_p/\delta e = 0$, where
\bqn
  \mathcal{L}_p &=& \frac{1}{2}\left(\frac{N^2}{e}\dot{t}^2 + e\Big[F(\zeta) - 2\zeta F'(\zeta)\Big]\right)\nb\\
	& =& \frac{1}{2}\left(\frac{N^2}{e}\dot{t}^2 - e\left[\zeta + \frac{3\zeta^2}{M_A^2}\right]\right).
\eqn
Thus, 
we will first calculate   $\delta \zeta/\delta e$,  implicitly by applying $\delta/\delta e$ to both sides of (\ref{zetaeq}),
which yields,
\bq
\frac{\delta \zeta}{\delta e} 
 = -\frac{2M_A^4\dot{r}^2}{e^3 f(M_A^4 + 8M_A^2 \zeta + 12 \zeta^2)}.
 \eq
Substituting this into the expression
$$
\frac{\delta  \mathcal{L}_p }{\delta e} = \frac{1}{2}\left(-\frac{N^2}{e^2}\dot{t}^2 - \left[\zeta + \frac{3\zeta^2}{M_A^2}\right]
- e\left[\frac{\delta \zeta}{\delta e} + \frac{6\zeta}{M_A^2}\frac{\delta \zeta}{\delta e}\right]\right),
$$
we find the following analog of equation (\ref{eqq6}),
\bqn
\label{4.6analog}  
&&  \zeta  \left(e^2 M_A^2+2 N^2 \dot{t}^2\right)+5 e^2 \zeta ^2+\frac{6 e^2 \zeta^3}{M_A^2} \nb\\
& & ~~~~~~~~~~~~ + M_A^2
   \left(N^2 \dot{t}^2- \frac{2 \dot{r}^2}{f}\right)  = 0,
\eqn
where $\zeta$ is given implicitly by Eq.(\ref{zetaeq}). Note that in the limit $M_A \rightarrow \infty$,  the above equation   reduces 
precisely to  Eq.(\ref{eqq6}) for $F(\zeta) = \zeta$ and $N^r = 0$,  as expected.

On the other hand, the analog of Eq.(\ref{eqq8}) is simply
\bq
\label{4.7analog}  
  N^2 \dot{t} = eE.
\eq

Using Eqs.(\ref{zetaeq}) and (\ref{4.7analog}) to eliminate $\dot{r}$ and $\dot{t}$ from Eq.(\ref{4.6analog}), we find \footnote{In the limit $M_A \to \infty$, this equation reduces to 
$\zeta - \frac{E^2}{N^2} = 0,$
which is again consistent with the case $F(\zeta) = \zeta$.}
$$
\frac{\delta  \mathcal{L}_p }{\delta e} = \frac{1}{2}\left(\zeta + \frac{\zeta^2}{M_A^2} - \frac{E^2}{N^2}\right) = 0.
$$
Solving this equation for $\zeta$, we infer that  
$$
\zeta = -\frac{M_A^2}{2} + \frac{M_A}{2N}\sqrt{4 E^2 + M_A^2 N^2}.
$$
Substitution of this expression    into Eq.(\ref{4.6analog}) yields
\bqn
\label{4.6analog2} 
& &  \frac{M_A \left(N^2 \left(\frac{2
   \dot{r}^2}{e^2 f}+M_A^2\right)+4
   E^2\right)}{N \sqrt{4 E^2+M_A^2
   N^2}}-\frac{N^2 \dot{t}^2}{e^2}\nb\\
   & & ~~~~~~~~~ -\frac{3
   E^2}{N^2}-M_A^2  = 0.
\eqn
Replacing   $e$ by $N^2 \dot{t}/E$ and then solving the resulting equation for $\dot{r}/\dot{t}$, we find \footnote{In the limit� $M_A \to \infty$, this equation becomes 
$\frac{\dot{r}^2}{\dot{t}^2} = fN^2,$
which is again consistent with the case $F(\zeta) = \zeta$. } 
$$\frac{\dot{r}^2}{\dot{t}^2} 
= \frac{f N \left(4 E^2+M_A^2 N^2\right) \left(\sqrt{4 E^2+M_A^2 N^2}-M_A
   N\right)}{2 E^2 M_A}.$$
Thus, the trajectory  is given by
\bq
t = t_0 + \int \frac{dr}{H(r,E)},
\eq
where
\bqn
H(r,E) &=& \sqrt{\frac{f N \left(4 E^2+M_A^2 N^2\right)}{2 E^2 M_A}}\nb\\
& &  \times \sqrt{\sqrt{4 E^2+M_A^2 N^2}-M_A
   N}.
 \eqn

As an example, let us consider the Schwarzschild solution,  $N^2 = f = 1 - r_g/{r}$, for which we find
$$
H = 2\sqrt{\frac{E}{M_A r_g^{3/2}}}(r - r_g)^{3/4} + {\cal{O}}\Big((r- r_g)^{5/4}\Big), 
$$
as $r \to r_g$, so that $t$ remains finite.  On the other hand, as $M_A \to \infty$,
$$
H = \frac{r - r_g}{r_g} + \frac{3E^2}{2M_A^2} +  {\cal{O}}\left(\frac{1}{M_A^4}\right).
$$
Thus, if we take the limit $M_A \to \infty$ before letting the trajectory  approach $r_g$, then $t$ will blow up logarithmically as $r \to r_g$.
As a result,  a horizon exists in this limit. 

More generally, if $f$ has an $a$th order zero and $N^2$ has a $b$th order zero at $r = r_g$, as given in Eq.(\ref{fN}),
then, we find that
$$
H = 2\sqrt{\frac{Ef_0(r_g)N_0(r_g)}{M_A}} (r - r_g)^{\frac{a}{2} + \frac{b}{4}} +  {\cal{O}}\left((r-r_g)^{\frac{a}{2} + \frac{3b}{4}}\right), 
$$
as $r \to r_g$.  It follows that
\bqn
t &\simeq& t_0 + \frac{1}{2\sqrt{\frac{Ef_0(r_g)N_0(r_g)}{M_A}}} \nb\\
& & \times \cases{ \frac{(r-r_g)^{1 - \frac{a}{2} - \frac{b}{4}}}{1 - \frac{a}{2} - \frac{b}{4}},
 & $\frac{a}{2} + \frac{b}{4} \neq 1$, \cr
\ln(r - r_g), 
& $\frac{a}{2} + \frac{b}{4} = 1$.\cr}
\eqn
Therefore, $t$ blows up as $r \to r_g$,  if and only if
\bq
a + \frac{b}{2}  \geq 2,
\eq
which is exactly Eq.(\ref{eqq19}) for $n = 2$, as expected.

 \section{Vacuum Solutions with $N^{r} = 0$ }

\renewcommand{\theequation}{4.\arabic{equation}} \setcounter{equation}{0}
 
 When $N^{r} = 0$, the vacuum equations with  $J^t = v = p_{r} = p_{\theta} = J_{A} = J_{\varphi} = 0$
 yield the following most general solutions \cite{GSW},  
 \bq
 \lb{4.1}
  f(r) = 1 + \frac{C}{r} -\frac{1}{3} \Lambda_g r^2, \;\;\; 
  N = 1,\;\;\; N^{r} = 0 = \varphi, 
 \eq
with the Hamiltonian constraint
  \bq 
 \lb{4.3c}
\int{{\cal{L}}_{V}  e^{\nu} r^{2} dr} = 0,
\eq
where ${\cal{L}}_{V} = {\cal{L}}_{V}(r, \Lambda_{g}, C, g_{s})$, as defined in Eq.(\ref{2.5a}).


The gauge field  $A$ must satisfy the equations, 
\bqn
 \lb{4.3a}
& &  A' + A \nu' + \frac{1}{2} r F_{rr}= 0,\\
  \lb{4.3b}
& & r^{2}\big(A'' - \nu'A'\big) + r \big(A' + \nu' A\big) - A\big(1 - e^{2\nu}\big) \nb\\
& &~~~~~~~~~~~~~~~~~~~ +    e^{2\nu} F_{\theta \theta} = 0,
 \eqn
where $F_{ij}$ is  given by Eqs.(\ref{eq3a}) and (\ref{A.2}). Then, from  Eq.(\ref{4.3a}) we find that
\bq
\lb{4.4}
A = A_{0} e^{-\nu}  - \frac{1}{2}e^{-\nu}\int^{r}{r'e^{\nu(r')} F_{rr}(r') dr'},
\eq
where $A_{0}$ is an integration constant. The solutions with $\Lambda_{g} = 0$ were first studied in \cite{HMT,AP}. 

Since now we have $N = 1$ and $ b = 0$,  Eq.(\ref{eqq19}) shows that   a horizon exists only when $a \ge 2$.
It can be shown that for the solutions given by Eq.(\ref{4.1}), this is impossible for any 
 chosen $C$ and $\Lambda_{g}$. Therefore, it is concluded that {\em the solutions given by Eq.(\ref{4.1}) do not represent black holes}.
 
However, in some cases $f(r) = 0$ does have a real and positive root. So, there indeed exists some kind of coordinate singularities, 
and to obtain a maximally (geodesically) complete spacetime \footnote{Because of the breaking of the general covariance and the restricted
diffeomorphism (\ref{1.2}), it is not clear if this requirement
is still applicable  here in the HL theory. Even if it is not, some kind of extensions still seems needed.},  
some kind of extensions  are needed. Such extensions are also needed in order to determine the range
of $r$, from  which  the Hamiltonian constraint (\ref{4.3c}) can be carried out. Once this constraint is satisfied, one can integrate Eq.(\ref{4.4}) to obtain the gauge field
$A$. To this end,  we divide the   solutions into the
cases:
$(i)\; C  = \Lambda_{g} = 0,\;
(ii) \; C \not= 0,\; \Lambda_{g} = 0, \; 
(iii) \; C = 0,\; \Lambda_{g} \not= 0$, and 
$(iv) \; C \not= 0,\;   \Lambda_{g} \not= 0$.
The first case is trivial, and it corresponds to the Minkowski spacetime with
$\nu = \Lambda = 0$ and $A = A_{0}$. Thus, in the following we shall consider only the last three cases.

\subsection{$C \not= 0,\;\;\; \Lambda_{g} = 0$}

In this case the metric takes the form
\bq
\lb{4.5a}
ds^{2} = - dt^{2} + \frac{dr^{2}}{1 + \frac{C}{r}} + r^{2}d^{2}\Omega,
\eq
from which we find that
\bqn
\lb{4.5}
{\cal{L}}_{V} &=& 2\Lambda + \frac{3g_{3}C^2}{2\zeta^2 r^{6}}  +  \frac{3g_{6}C^3}{4\zeta^4 r^{9}}\nb\\
& & ~
+  \frac{45g_{8}C^2}{2\zeta^4 r^{8}}\left(1 + \frac{C}{r}\right),
\eqn
where $\Lambda = g_{0}\zeta^2/2$.
To consider the Hamiltonian constraint (\ref{4.3c}), we need to further distinguish the cases $C > 0$ and $C < 0$.

\subsubsection{$C > 0$}

When $ C > 0$, the metric (\ref{4.5a}) is singular only at $r = 0$, so the solution covers the whole spacetime
 $r \in (0, \infty)$. The singularity at the center  is a curvature one  \cite{CW}, as it can be seen  from the  expressions,
 \bqn
 \lb{4.5b}
 R^{ij}R_{ij} &=& \frac{3C^{2}}{2r^{6}},\nb\\
R^{i}_{j}R^{j}_{k}R^{k}_{i} &=& - \frac{3C^{3}}{4r^{9}}, \nb\\
\left(\nabla_{i}R_{jk}\right) \left(\nabla^{i}R^{jk}\right) &= &   \frac{45 C^2}{ 2 r^{8}}\left(1 + \frac{C}{r}\right).
\eqn
 Since   event horizons do not exist for $C> 0$, this singularity is also  naked.  Inserting it into Eq.(\ref{4.3c}), 
 we find that the Hamiltonian constraint is satisfied only when
\bq
\lb{4.6}
\Lambda = g_{3} = g_{6} = g_{8} = 0.
\eq
Considering Eq.(\ref{A.2}), we find that $F_{ij}$ now has only two non-vanishing terms, given by
\bq
\lb{4.7}
F_{ij} = -\left(F_{1}\right)_{ij} + \frac{g_{5}}{\zeta^{4}}\left(F_{5}\right)_{ij}.
\eq
Substituting it into Eqs.(\ref{4.3a}) and (\ref{4.3b}), we obtain 
\bq
\lb{4.8}
A = 1 + A_{0}\sqrt{1 + \frac{C}{r}}, \;\;\; g_{5} = 0.
\eq
It should be noted that the above solution holds not only   in the infrared (IR) regime but also in the UV.

\begin{figure}[tbp]
\centering
\includegraphics[width=8cm]{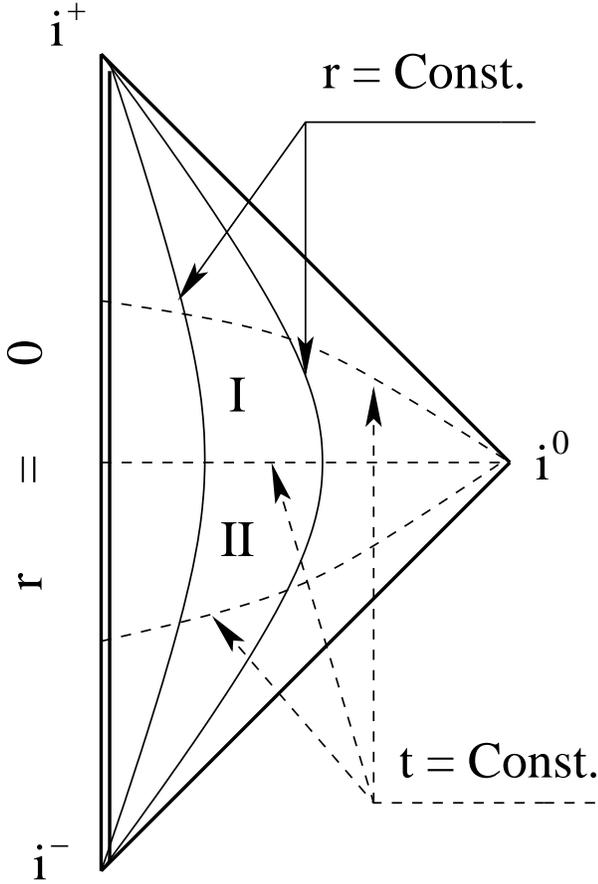}
\caption{The Penrose diagram for $N^{r} = 0,\; C > 0$ and $\Lambda_{g} = 0$. The double vertical solid lines represent the center ($r = 0$), at which the
spacetime is singular.  This singularity is clearly naked. Note that the restricted diffeomorphisims (\ref{1.2}) do not allow for the transformations
needed in order to draw Penrose diagrams. Therefore, these diagrams cannot be used to study the global structures of spacetimes in the 
HL theory but are included only for comparison.   }
\label{fig1}
\end{figure}

To study the global structure of the spacetime, let us first introduce a new radial coordinate $r^{*}$ via the relation
\bqn
\lb{4.8a}
r^{*} &\equiv& \int{\frac{dr}{\sqrt{1 + \frac{C}{r}}}}  =  - \frac{C}{2}\ln\frac{\left(\sqrt{r + C} + \sqrt{r}\right)^{2}}{C} \nb\\
& &  + \sqrt{r(r+C)}   
= \cases{0, & $ r = 0$,\cr
\infty, &$r = \infty$.}
\eqn
In terms of $r^{*}$ the metric takes the form,
\bq
\lb{4.8b}
ds^{2} = - dt^{2} + { dr^{*}}^{2} + r^{2}(r^{*})d^{2}\Omega.
\eq
Then, one might introduce the two double null coordinates $u$ and $v$ via the relations,
\bq
\lb{4.8c}
u = \tan^{-1}(t + r^{*}),\;\;\;
v = \tan^{-1}(t - r^{*}),
\eq
so that the metric finally takes the form, 
\bq
\lb{4.8d}
ds^{2} = -\frac{dudv}{\cos^{2}u\cos^{2}v} + r^{2}(u, v)d^{2}\Omega,
\eq
where $- \pi/2 \le u, v \le \pi/2$. The corresponding Penrose diagram is given by Fig. \ref{fig1}.

However, the coordinate transformations (\ref{4.8c}) are not allowed by the   foliation-preserving diffeomorphisms Diff($M, \; {\cal{F}}$) of Eq.(\ref{1.2}). So, 
in the HL theory the restricted  diffeomorphisms do not permit    Penrose diagrams.  In addition, due to the breaking of the general covariance, even if
one were allowed to do so, the causal structure of the spacetime cannot be studied in terms of it, as shown explicitly  in the previous sections for 
the Newtonian  theory.  

\begin{figure}[tbp]
\centering
\includegraphics[width=8cm]{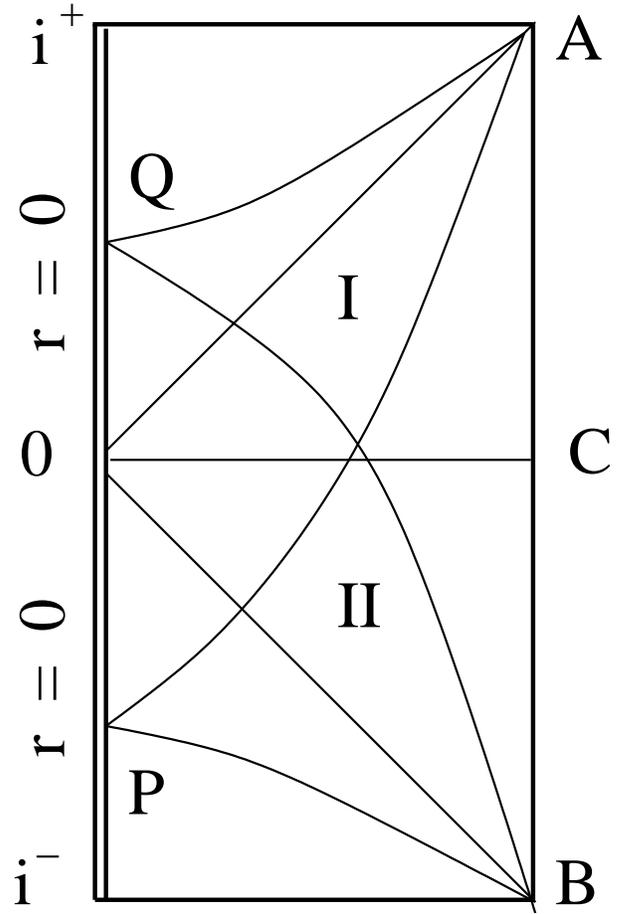}
\caption{The global structure of the spacetime in the ($\bar{t},\; \bar{r}^{*}$)-plane  for $N^{r} = 0,\; C > 0$ and $\Lambda_{g} = 0$. 
The double vertical solid lines represent the center ($r = 0$), at which the
spacetime is singular.   The vertical line $AB$ represents the spatial infinity  $r = \infty$, while the horizontal line $i^{+}A\; (i^{-}B)$ 
is the line where $t = \infty\; (t = -\infty)$. The lines $t = $ Constant are the straight lines parallel to $OC$, while the ones $r = $ Constant 
are the straight lines parallel to $i^{-}i^{+}$. The lines $BP, BO, BQ, PA, OA$ and $QA$ represent the radial null geodesics.}
\label{fig1a}
\end{figure}

Allowed are the coordinate transformations 
\bq
\lb{4.8e}
t = \tan{\bar{t}},\;\;\;
r^{*} = \tan{\bar{r}^{*}},
\eq
where  $- \pi/2 \le \bar{t} \le \pi/2$ and  $0 \le  \bar{r}^{*} \le \pi/2$. Then, the global structure of the spacetime is given by Fig. \ref{fig1a}.
 
\subsubsection{$ C < 0$}

In this case, setting $C = - 2M < 0$, the corresponding metric reads, 
\bq
\lb{4.8f}
ds^{2} = - dt^{2} + \left(1 - \frac{2M}{r}\right)^{-1}dr^{2} + r^{2}d^{2}\Omega.
\eq
This is the solution first found in \cite{HMT}, in which it was
argued that the relativistic lapse function  should be ${\cal{N}} = N - A$  in the IR.  It is not clear how to then relate ${\cal{N}}$ to $N$ and $A$ in other regimes.   Instead, 
in this paper we shall simply take the point of view that $A$ and $\varphi$ are just gravitational gauge fields, and their effects on the 
spacetime itself occur only through  the field equations \cite{GSW}. 
With the above arguments,  we  can consider the solution valid in any regimes, including   the IR and UV.  

Let us first note that
the metric (\ref{4.5a}) is asymptotically flat and singular at both $r = 0$ and $ r =  2M$. The singularity at $r = 0$ is a
curvature one,
as can be seen from Eq.(\ref{4.5b}), but the one at $r =  2M$ is more peculiar. In particular,  in the region $r < 2M$ both $t$ and $r$ 
are timelike, in contrast to GR where $t$ and $r$ exchange their roles across $r = 2M$.  
All the above indicate that the nature of the singularity at $r = 2M$ now is different. In fact, as to be shown explicitly below, the region $r < 2M$ actually is not part of the 
spacetime. 

To see this closely,    let us 
first consider the radial timelike geodesics. It can be shown that they are given by,
\bqn
\lb{4.9}
t &=&E \tau + t_{0},\nb\\
\tau  &=& \pm \frac{1}{\sqrt{E^{2} -1}}\Bigg\{M \ln\Big[(r -M) + \sqrt{r(r -2M)}\Big] \nb\\
& & ~~~~~~~~~~~~~~~~~ -  \sqrt{r(r -2M)}\Bigg\} +  \tau_{0},
\eqn
where $E$ is an integration constant, and $\tau$ denotes the proper time. 
The constant $\tau_{0}$ is chosen so that $\tau(r_{0}) = 0$ at the initial position of the test particle, $r = r_{0} > 2M$. 
The ``+" (``-") sign corresponds to the out-going (in-going) radial geodesics. It is clear that, starting at any given finite radius $r_{0}$,
 observers that follow the null geodesics will arrive at $r = 2M$ within a finite proper time \footnote{As shown in the last section, massless test
 particles in the HL theory do not follow  null geodesics, because of the non-relativistic dispersion relations (\ref{1.4}). In other words,  in the HL theory
 particles that follow the null geodesics are not massless and even may not be   test particles.}.  Setting
\bq
\lb{4.10}
e^{\alpha}_{(0)} \equiv \frac{dx^{\alpha}}{d\tau} = \left(E, - \sqrt{(E^{2} -1)f}, 0, 0\right),
\eq
where $f \equiv 1 - 2M/r$, we find that the  spacelike unit vectors, 
\bqn
\lb{4.11}
e^{\alpha}_{(1)}   &=& \left(\sqrt{E^{2} -1}, - E \sqrt{f}, 0, 0\right),\nb\\
e^{\alpha}_{(2)}   &=&  \frac{1}{r}\left(0, 0, 1, 0\right),\nb\\
e^{\alpha}_{(3)}   &=& \frac{1}{r\sin\theta} \left(0, 0, 0, 1\right),
\eqn
together with $e^{\alpha}_{(0)}$ form a freely-falling frame,
\bq
\lb{4.12}
e^{\alpha}_{(a)} e_{\alpha\; (b)}   = \eta_{ab}, \;\;\;
e^{\alpha}_{(0)}D_{\alpha} e^{\beta}_{(a)} = 0, 
\eq
where $D_{\alpha}$ denotes the 4D covariant derivatives, and $\eta_{ab}$ is the 4D Minkowski metric
with $a, b, = 0, ..., 3$. Then, from the geodesic deviations,
\bq
\lb{4.13}
\frac{D^{2}\eta^{a}}{D\tau^{2}} + {\cal{K}}^{a}_{b}\eta^{b} = 0,
\eq
where ${\cal{K}}_{ab} \equiv - R_{\sigma\alpha\beta\gamma} e^{\sigma}_{(a)}  e^{\alpha}_{(0)}e^{\beta}_{(0)}
e^{\gamma}_{(b)}$ denotes the tidal forces exerting on the observers, we find that  in the present case 
${\cal{K}}_{ab} $ is given by
\bq
\lb{4.14}
{\cal{K}}_{ab} = - \frac{(E^{2} -1)M}{r^{3}}\left(\delta^{2}_{a}\delta^{2}_{b} + \delta^{2}_{a}\delta^{3}_{b}\right).
\eq
Clearly, ${\cal{K}}_{ab}$ is finite at $r = 2M$. All the above considerations indicate that the singularity at $r = 2M$ is a coordinate
one, and to have a (geodesically) complete spacetime, extension beyond this surface is needed. However, unlike
that in GR,  any extension must be restricted to the Diff($M, \; {\cal{F}}$) of Eq.(\ref{1.2}). Otherwise,
the resulting solutions do not satisfy the field equations. Explicit examples of this kind were given in \cite{CW}. 

\begin{figure}[tbp]
\centering
\includegraphics[width=8cm]{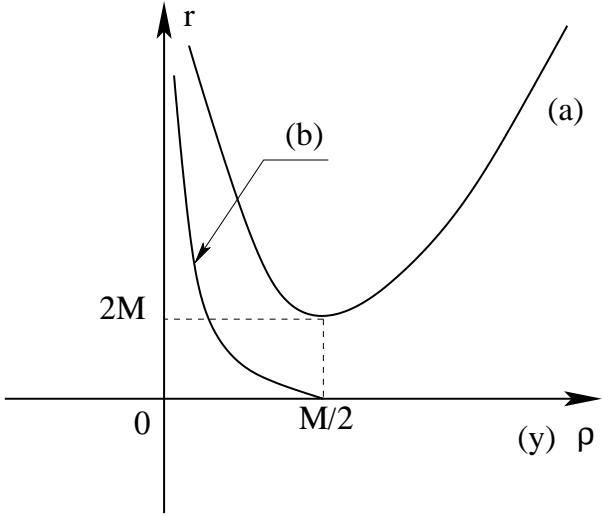}
\caption{The function $r$ defined: (a)  by Eq.(\ref{4.15}); and (b)  by Eq.(\ref{4.16}). } 
\label{fig2}
\end{figure}

In \cite{HMT}, the isotropic coordinate $\rho$ was introduced, 
\bq
\lb{4.15}
r = \rho\left(1 + \frac{M}{2\rho}\right)^{2},
\eq
in terms of which  
 the metric (\ref{4.5a}) takes the form,
 \bq
\lb{4.5c}
ds^{2} = - dt^{2} + \left(1 + \frac{M}{2\rho}\right)^{4}\Big(d\rho^{2} + \rho^{2}d^{2}\Omega\Big),
\eq 
which is non-singular for $\rho > 0$. However, this cannot be considered as an extension to the region $r < 2M$, 
as now the geometrical radius $r$ is still restricted to $r \in (2M, \infty)$ for $\rho > 0$, as shown by Curve (a)  in Fig. \ref{fig2}. 
Instead, it connects two asymptotic regions, where $r = 2M$ acts as a throat, a situation  quite similar
to the Einstein-Rosen bridge \cite{MTW}.  However, a fundamental difference of the metric (\ref{4.5c}) from the corresponding one 
in GR  is that  it is not singular for 
any $\rho \in (0, \infty)$, while in GR the metric still has a coordinate singularity at $\rho = M/2$ (or $r = 2M$) 
\cite{MTW}.  Therefore, in the HL theory Eq.(\ref{4.5c}) already represents   an extension of the metric (\ref{4.5a}) 
 beyond the surface $r = 2M$. Since this extension is   analytical, it is  unique. It is remarkable to note that in this
extension the metric has the correct signature.  

It should be noted that  the Einstein-Rosen bridge is not stable in GR  \cite{MTW}.
Therefore, it would be  very interesting to know if this is still the case  in the HL theory.

\begin{figure}[tbp]
\centering
\includegraphics[width=8cm]{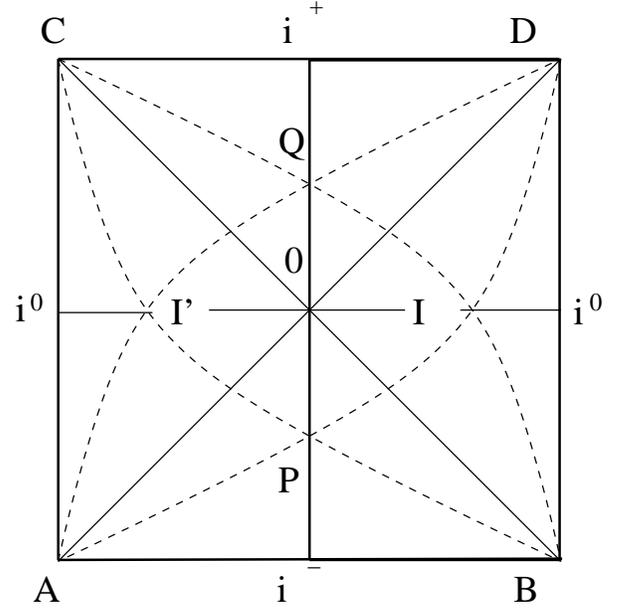}
\caption{The global structure of the spacetime  for $N^{r} = 0, \; C = - 2M <  0$ and $\Lambda_{g} = 0$.    The vertical line $i^{+}i^{-}$ represents the Einstein-Rosen throat 
($r = r_{g} \equiv 2M$), which is non-singular and connects the two asymptotically-flat  regions $I$ and $I'$.  The horizontal line $AB\; (CD)$ is the line where $t = -\infty \; (\infty)$, while the 
vertical lines $CA$ and $DB$ are the lines where $r = \infty$. The lines $t = $ Constant are the straight lines parallel to $i^{0}i^{0}$, 
while the ones $r = $ Constant are the straight lines parallel to $i^{-}i^{+}$.  The curved dotted lines $AD$ and $BC$, as well as the solid straight lines $AD$ and $BC$,  are the radial null geodesics.}
\label{fig3a}
\end{figure}

To study its global structure, we introduce the coordinate $r^{*}$ by
\bqn
\lb{4.5ca}
r^{*} &\equiv&  \int{\left(1 + \frac{M}{2\rho}\right)^{2}d\rho} = M\ln\left(\frac{2\rho}{M}\right)\nb\\
& & + \rho\left(1- \frac{M^{2}}{4\rho^{2}}\right) = \cases{-\infty, & $\rho = 0$,\cr
\infty, & $\rho = \infty$.\cr}
\eqn
Then, in terms of $r^{*}$  the metric can be also cast in the form of Eq.(\ref{4.8b}). Following what was done in that case, one can see that the global structure of the 
spacetime is   given by Fig. \ref{fig3a}. 

\begin{figure}[tbp]
\centering
\includegraphics[width=8cm]{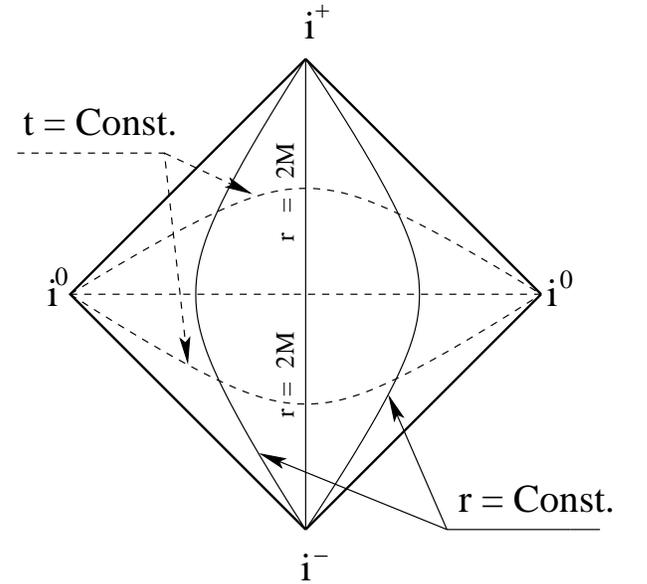}
\caption{The Penrose diagram for $N^{r} = 0, \; C = - 2M <  0$ and $\Lambda_{g} = 0$.  The straight lines $i^{+}i^{0}$ represent the future null infinities
at which we have $r = \infty$ and $t = \infty$, while the ones  $i^{-}i^{0}$ represent the past null infinities where
$r = \infty$ and $t = - \infty$. The vertical line $i^{+}i^{-}$ represents the Einstein-Rosen throat ($r = 2M$), which is non-singular and connects the 
two asymptotically-flat
 regions. }
\label{fig3}
\end{figure}

To compare it with that  given in GR,  the corresponding Penrose diagram is presented  in Fig. \ref{fig3}, although it is forbidden in the HL theory
by the  foliation-preserving diffeomorphisms Diff($M, \; {\cal{F}}$) of Eq.(\ref{1.2}), as mentioned above.

It is interesting to see which kind of matter fields can give rise to such a spacetime in GR. To this purpose, we first
calculate the corresponding 4-dimensional Einstein tensor, 
\bq
\lb{4.17b}
{}^{(4)}G_{\mu\nu} = \frac{2M}{r^{3} f} \delta^{r}_{\mu} \delta^{r}_{\nu}  - \frac{M}{r}\left( \delta^{\theta}_{\mu} \delta^{\theta}_{\nu}
+ \sin^{2}\theta \delta^{\phi}_{\mu} \delta^{\phi}_{\nu} \right), 
\eq
which  corresponds to an anisotropic fluid, $T^{GR}_{\mu\nu} = \rho^{GR}u_{\mu}u_{\nu} + p^{GR}_{r}r_{\mu}r_{\nu}
+ p^{GR}_{\theta}\left(\delta^{\theta}_{\mu} \delta^{\theta}_{\nu}
+ \sin^{2}\theta \delta^{\phi}_{\mu} \delta^{\phi}_{\nu}\right)$,  with $\rho^{GR} = 0,\; p^{GR}_{r} = M/(4\pi Gr^{3})$ and $p^{GR}_{\theta} = - M r/(8\pi G)$,
where $u_{\mu} = \delta^{t}_{\mu}$ and $r_{\mu} = f^{-1/2}\delta^{r}_{\mu}$. Clearly,
such a fluid does not satisfy any of the energy conditions \cite{HE73}. In particular, when $r \gg 1$  the tangential pressure becomes unbounded 
from below, while the radial pressure vanishes. Such a fluid is usually considered as non-physical in GR. However, in the current setup the spacetime 
is vacuum, 
and one cannot eliminate it by simply considering the energy conditions. Then, if the configuration
is stable, one can use it to construct time-machines \cite{Visser}.

Inserting Eq.(\ref{4.5}) into Eq.(\ref{4.3c}), and considering  the fact that the range of  $r$ now is  $ r \in (2M, \infty)$,
 we find that the Hamiltonian constraint is satisfied, provided that
\bq
\lb{4.17}
 \Lambda = 0, \;\;\; 20\big(g_{6} - 3g_{8}\big) - {231}g_{3} \zeta^{2}M^{2} = 0.
 \eq
Then, Eqs.(\ref{4.3a})  and (\ref{4.3b}) have the solution,
\bqn
\lb{4.17a}
A &=& 1 + A_{0} \sqrt{1 - \frac{2M}{r}} + \frac{g_{3}}{40\zeta^{2}M^{2}r^{6}}\Big[16\big(r-M\big)r^{5} \nb\\
& & - 8M^{2}\big(r + M\big)r^{3} - 3M^{3}\big(5r^{2} + 7 Mr + 1050 M^{2}\big)\Big],\nb\\
g_{5} &=& g_{8} = 0.  
\eqn

It is interesting to note that, replacing $\rho$ by $-y$ we find that in terms of $y$  metric (\ref{4.5c})
 takes the form, 
 \bq
\lb{4.5d}
ds^{2} = - dt^{2} + \left(1 - \frac{M}{2y}\right)^{4}\Big(dy^{2} + y^{2}d^{2}\Omega\Big),
\eq
 from which we can see that the geometrical radius now is given by
\bq
\lb{4.16}
r = y\left(1 - \frac{M}{2y}\right)^{2}.
\eq
Clearly, the whole region $ 0 \le r < \infty$ now is mapped to $  0 < y \le M/2$, as shown by Curve (b) in Fig. \ref{fig2}.
Metric (\ref{4.5d}) can be also obtained from metric (\ref{4.5c}) by the replacement,
$M \rightarrow -M$ and $\rho \rightarrow
y$. So, it must correspond to the case $C > 0$, i.e., the one with a negative mass, described
in the previous sub-case.

\subsection{$C = 0,\;\;\; \Lambda_{g}  \not= 0$}

We have
\bq
\lb{4.18a}
\nu = - \frac{1}{2}\ln\left(1 - \frac{1}{3}\Lambda_{g} r^{2}\right),
\eq
for which we find that
\bqn
\lb{4.19a}
{\cal{L}}_{V} &=& 2\big(\Lambda - \Lambda_{g}\big) + \frac{4(3g_{2} + g_{3})}{3\zeta^{2}}\Lambda^{2}_{g}\nb\\
& &  + 
\frac{8(9g_{4} + 3g_{5} + g_{6})}{9\zeta^{4}}\Lambda^{3}_{g},\nb\\
F_{ij} &=& \frac{g_{ij}}{9\zeta^{4}}\Big[3\zeta^{4}\big(\Lambda_{g} - 3\Lambda\big) + 2\zeta^{2}\big(3g_{2} + g_{3}\big)\Lambda_{g}^{2}\nb\\
& & ~~~~~~~
+ 4\big(9g_{4} + 3g_{5}+ g_{6}\big)\Lambda_{g}^{3}\Big].
\eqn
To study the solutions further, we consider the cases $\Lambda_{g} > 0 $ and  $\Lambda_{g} < 0$, separately.

\subsubsection{$\; \Lambda_{g} < 0$}

In this case, defining 
$r_{g} \equiv \sqrt{{3}/{\left|\Lambda_{g}\right|}}$,
we find that the corresponding metric takes the form,
\bq
\lb{4.21}
ds^{2} = - d{t}^{2} + \frac{dr^{2}}{1 + \left(\frac{r}{r_{g}}\right)^{2}} + r^{2} d^{2}\Omega,
\eq
which shows  that the metric is not singular except  at $r = 0$. 
But, it  can be shown
that this is a coordinate singularity. Setting
\bqn
\lb{4.22}
r^{*} &\equiv& \int{\frac{dr}{\sqrt{1 +  \left(\frac{r}{r_{g}}\right)^{2}}}}\nb\\
&=& r_{g}\ln\left\{\frac{r}{r_{g}} + \sqrt{1 +  \left(\frac{r}{r_{g}}\right)^{2}}\right\},
\eqn
one can cast the metric (\ref{4.21}) exactly in the form of Eq.(\ref{4.8b}). Then, its global structure is that of Fig. \ref{fig1a}, and 
the corresponding Penrose diagram is  given by Fig. \ref{fig1}, but now   the center $r = 0$ is free of any spacetime singularity. 
Thus, the range of $r$ now is  $r \in [0, \; \infty)$. 
We then find that the Hamiltonian constraint (\ref{4.3c})  is satisfied, provided that ${\cal{L}}_{V} = 0$, i.e., 
\bqn
\lb{4.22a}
& & \Lambda \zeta^{4}r_{g}^{6} +   6\big(3g_{2} + g_{3}\big) \zeta^{2}r_{g}^{2}  - 12\big(9g_{4} + 3g_{5}  + g_{6}\big) \nb\\
& & ~~~~~~~~~~~~~~~~~~~~~~~~~~~~~~~~~~~~~ =  -3\zeta^{4}r_{g}^{4}. 
\eqn
Inserting the above into   Eqs.(\ref{4.3a})  and (\ref{4.3b}), we obtain the solution, 
\bq
\lb{4.22b}
A = A_{0}\sqrt{1 + \left(\frac{r}{r_{g}}\right)^{2}} + A_{1},
\eq
where $A_{1}$ is a constant, given by
\bq
\lb{4.22c}
A_{1} \equiv  1- \Lambda r_{g}^{2} - \frac{3 - 3g_{2} - g_{3}}{\zeta^{2}r^{2}_{g}}.
\eq


\subsubsection{$\; \Lambda_{g} > 0$}

\begin{figure}[tbp]
\centering
\includegraphics[width=8cm]{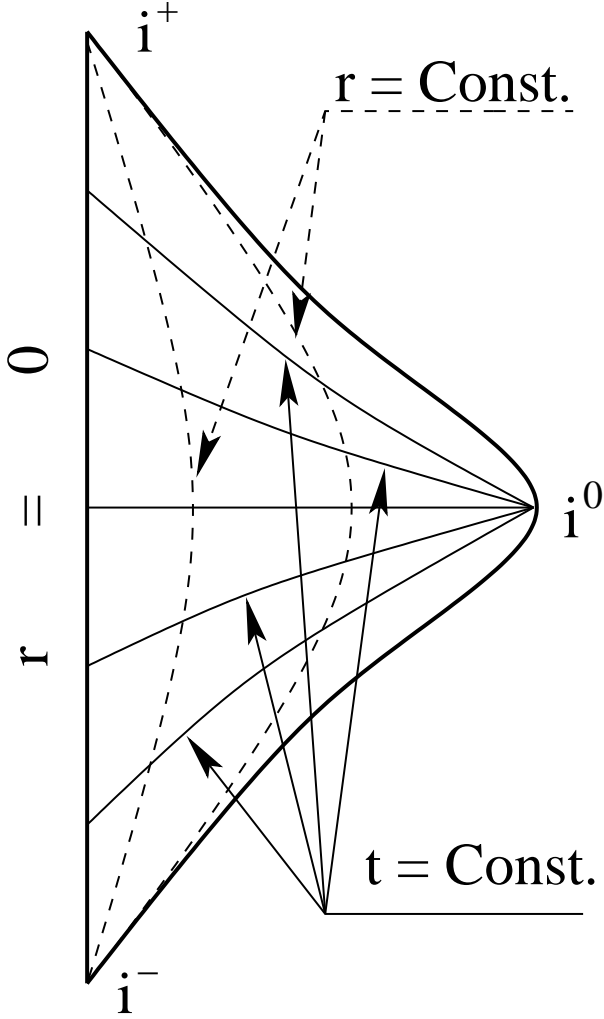}
\caption{The Penrose diagram for $N^{r} = 0,\; C =0$ and $\Lambda_{g} > 0$, which is the Einstein static universe.    The curves $i^{-}i^{0}$ and
$i^{+}i^{0}$ are, respectively, the lines where $t = -\infty,\; x = \pi$, and $t = + \infty,\; x = \pi$.}
\label{fig5}
\end{figure}

In this case,  the corresponding metric takes the form,
\bq
\lb{4.23}
ds^{2} = - dt^{2} + \frac{dr^{2}}{1 - \left(\frac{r}{r_{g}}\right)^{2}} + r^{2} d^{2}\Omega.
\eq
Clearly,  the metric has wrong signature in the region $r > r_{g}$. In fact, the hypersurface $r = r_{g}$ already represents the geometrical boundary
of the spacetime, and any extension beyond it is not needed. To see this clearly, we first introduce the coordinate $r^{*}$ via the relation,
\bq
\lb{4.24}
r^{*} \equiv \int{\frac{dr}{\sqrt{1 -  \left(\frac{r}{r_{g}}\right)^{2}}}}
= r_{g}\arcsin\left(\frac{r}{r_{g}}\right).
\eq
Then, in terms of $r^{*}$ the corresponding metric can be cast in the form $ds^{2} = r_{g}^{2}d\bar{s}^{2}$, where
\bq
\lb{4.25}
d\bar{s}^{2} = - d\bar{t}^{2} + dx^{2} +  \sin^{2}x d^{2}\Omega,
\eq
with $\bar{t} = t/r_{g},\; x = r^{*}/r_{g}$. But, this is exactly the homogeneous and isotropic Einstein static universe, which is geodesically    complete  for
$ - \infty < \bar{t} < \infty,\; 0 \le x \le \pi,\;
0 \le \theta \le \pi$ and $0 \le \phi \le 2\pi$,  with an $R\times{S^{3}}$ topology \cite{HE73}. Then, it is easy to see that its global structure is given by Fig. \ref{fig1a}, but
now the vertical line $i^{-}i^{+}$ is free of spacetime singularity, and the line $AB$ is the one where $r = r_{g}$ (or $x = \pi$). The corresponding Penrose diagram is given by Fig. \ref{fig5}.

Therefore, in this case the range of $r$ is $r \in [0, \; r_{g}]$. Then, the Hamiltonian constraint (\ref{4.3c})
requires,
\bqn
\lb{4.26}
& & \Lambda \zeta^{4}r_{g}^{6} +   6\big(3g_{2} + g_{3}\big) \zeta^{2}r_{g}^{2}  +12\big(9g_{4} + 3g_{5}  + g_{6}\big) \nb\\
& & ~~~~~~~~~~~~~~~~~~~~~~~~~~~~~~~~~~~~~ =  3\zeta^{4}r_{g}^{4}. 
\eqn
Hence,  Eqs.(\ref{4.3a})  and (\ref{4.3b}) have the solution,
\bq
\lb{4.27}
A = A_{0}\sqrt{1 - \left(\frac{r}{r_{g}}\right)^{2}} + A_{2},
\eq
where $A_{2}$ is another integration constant, given by
\bq
\lb{4.28}
A_{2} \equiv  1+ \Lambda r_{g}^{2} + \frac{3 - 3g_{2} - g_{3}}{\zeta^{2}r^{2}_{g}}.
\eq

It should be noted that in GR the Einstein static universe is obtained by the exact balance between the gravitational attraction of matter ($\rho_{m} = \rho_{c},\; p_{m} = 0$)
and the cosmic repulsion ($\Lambda = \Lambda_{c}$), where $\Lambda_{c} = 4\pi G \rho_{c}$. As a result, the configuration is not stable against small perturbations
\cite{Hart}. However, in the present case since the spacetime is vacuum, Eq.(\ref{4.26}) suggests that the balance is made by the attraction of the high-order curvature 
derivatives and the cosmic  repulsion, produced by both $\Lambda$ and $\Lambda_{g}$. Then, it would be very interesting to know whether it is stable or not in the current setup.

\subsection{$C \not= 0,\;\;\;  \Lambda_{g}  \not= 0$}

When $\Lambda_{g}, \ C \not= 0$, we find that
\bqn
\lb{4.29a}
 & & R^{ij}R_{ij} =  \frac{9 C^{2} + 8 \Lambda_{g}^{2} r^{6}}{6r^{6}},\nb\\
& & R^{i}_{j}R^{j}_{k}R^{k}_{i} = \frac{1}{36 r^{9}}\Big(27 C^{3} + 108 \Lambda_{g}C^{2} r^{3} + 32 \Lambda_{g}^{3}r^{9}\Big), \nb\\
& & \left(\nabla_{i}R_{jk}\right) \left(\nabla^{i}R^{jk}\right) =    \frac{45C^{2}}{2r^{8}}\Bigg(1 + \frac{C}{r} - \frac{1}{3}\Lambda_{g}r^{2}\Bigg),~~~~~~
\eqn
from which one can see that the spacetime is singular at $r = 0$. Moreover, we find from (\ref{A.2}) that 
\bqn
\label{Frrgeneral}
F_{rr}&= &\frac{1}{36 r^8 \zeta^4 F(r)}  \biggl\{
-27 C^3 (22 g_5+25 g_6-20 g_8) \nonumber\\
	\nonumber \\
&& -81 C^2 r (8 g_5+9 g_6-7 g_8 )
	 \nonumber\\
& & -9 C^2  r^3 \Bigl[\Lambda_g(-26 g_5-30 g_6+25 g_8) + \zeta ^2 g_3 \Bigr]
	 \nonumber\\
& & +12 C r^6 \Bigl[-3 \zeta ^4 +\Lambda_g \zeta ^2 (12 g_2+5 g_3)
	\nonumber \\
&&\qquad\qquad +\Lambda_g^2 (36 g_4+14 g_5+6g_6-g_8)\Bigr]
	 \nonumber\\
& &+4 r^9 \Bigl[-3 \zeta ^4 (3 \Lambda - \Lambda_g)+2 \zeta ^2 \Lambda_g^2 (3 g_2+g_3)
	\nonumber \\
&&\qquad\qquad +4 \Lambda_g^3 (9 g_4+3 g_5+g_6)\Bigr]\biggr\},
\eqn
where the third-order polynomial $F(r)$ is defined by 
$$F(r) = C + r - \frac{\Lambda_g}{3}r^3 \qquad \text{i.e.} \qquad e^{2\nu} = \frac{r}{F(r)}.$$
The function $\mathcal{L}_V$ is given by
\begin{equation}\label{LVintegrand}
  \mathcal{L}_V = \frac{\alpha + \beta r + \gamma r^3 + \delta r^9}{36 r^9 \zeta ^4},
  \end{equation}
where
\begin{eqnarray*}
  \alpha & =& 27 C^3 g_6+810 C^3 g_8,
  	\\
  \beta &=& 810 C^2 g_8,
  	\\
   \gamma &=& 108 C^2 g_5 \Lambda_g+108 C^2 g_6 \Lambda_g-270 C^2 g_8 \Lambda_g\nb\\
   & &  +54 C^2 g_3 \zeta ^2,
  	\\
   \delta &=& 144 g_2 \zeta ^2 \Lambda_g^2+288 g_4 \Lambda_g^3+96 g_5 \Lambda_g^3+32 g_6 \Lambda_g^3
	\\
&& +48 g_3 \zeta^2\Lambda_g^2+72 \zeta ^4 \Lambda -72 \zeta ^4 \Lambda_g.
\end{eqnarray*}

All the quantities in (\ref{4.29a}) are finite for any $r \neq 0$. On the other hand,
from Eq.(\ref{4.1}) one can see that the metric coefficient $g_{rr}$ could become singular at some points. 
To study the nature of these singularities, we distinguish the four cases, $C > 0,\; \Lambda_{g} > 0$;  $C > 0,\; \Lambda_{g} < 0$;  $C < 0,\; \Lambda_{g} > 0$; 
and $C < 0,\; \Lambda_{g} < 0$. 

\subsubsection{$C > 0,\; \Lambda_{g} > 0$}
\begin{figure}[tbp]
\centering
\includegraphics[width=8cm]{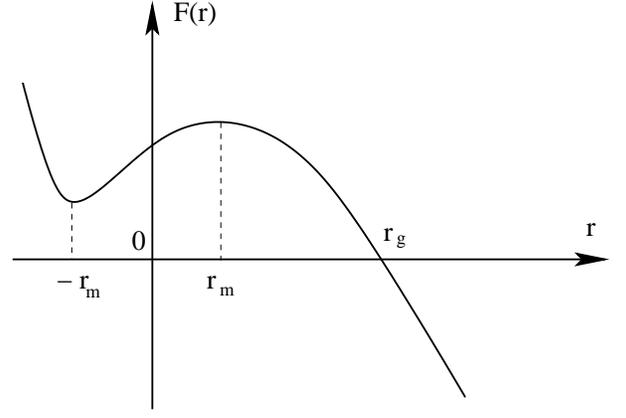}
\caption{ The function $F(r) \equiv re^{-2\nu}$ for $N^{r}  = 0,\; C > 0$ and $\Lambda_{g} > 0$, where $r_{m} = 1/\sqrt{\Lambda_{g}}$.}
\label{fig5a}
\end{figure}

\begin{figure}[tbp]
\centering
\includegraphics[width=8cm]{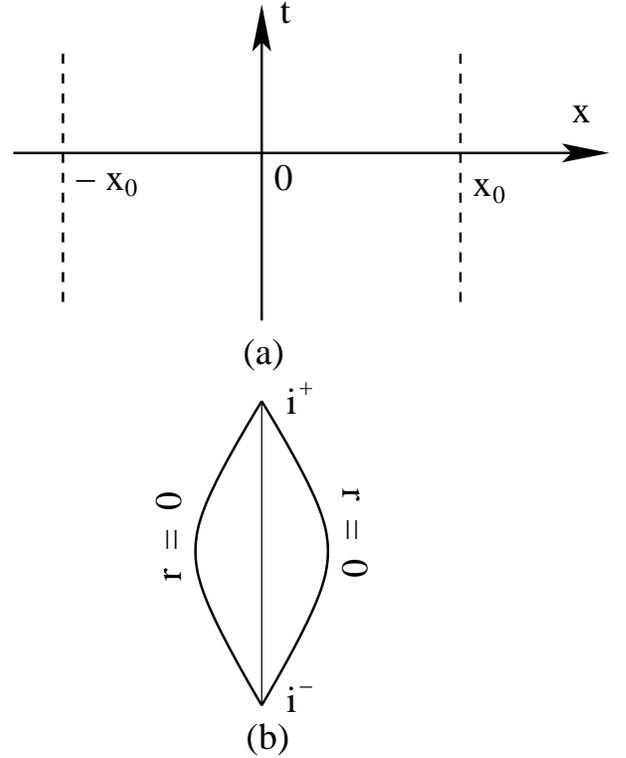}
\caption{(a) The spacetime in the ($t, x$)-plane, where $x_{0} \equiv \sqrt{r_{g}}$. (b) The Penrose diagram for $N^{r} = 0,\; C > 0 ,\; \Lambda_{g} > 0$.    The curves $i^{-}i^{+}$  
are the lines where  $r = 0$, at which the spacetime is singular. The straight line $i^{-}i^{+}$ represents the surface $r = r_{g}$.}
\label{fig6}
\end{figure}

In this case, the polynomial $F(r)$ has exactly one real positive root at, say, $r = r_{g}(C, \Lambda_{g})> 0$, as shown in Fig. \ref{fig5a}. We find that
\bq
\lb{4.30}
e^{2\nu} = \frac{r}{D(r)(r_{g} - r)},
\eq
where $D(r) \equiv \Lambda_{g}(r^{2} + r_{g} r + d)/3$, $d = r_{g}^{2} - 3/\Lambda_{g}$, and
$D(r) > 0$ for all $r > 0$.   Introducing the coordinate $x$ via the relation
\bq\label{xrrelation}
x = \int{\frac{dr}{2\sqrt{r_{g} - r}}} = - \sqrt{r_{g} - r}, 
\eq
or, inversely, $r = r_{g} - x^{2}$, 
the corresponding metric in terms of $x$ takes the form
\bq
\lb{4.31}
ds^{2} = - dt^{2} + \frac{4(r_{g} - x^{2})}{D(x)} d^{2}{x} +  \big(r_{g} - x^{2}\big)^{2} d^{2}\Omega, 
\eq
where $D(x) = \Lambda_{g}(x^{4} - 3r_{g}x^{2} + 3r_{g}^{2} - 3/\Lambda_{g})/3 > 0$ for $|x| < \sqrt{r_{g}}$. Clearly, the coordinate singularity at $r = r_{g}$ (or
$x = 0$) now is removed, and the metric is well defined for $|x| < \sqrt{r_{g}}$. At  the points, $ x = \pm  \sqrt{r_{g}}$ (or $r = 0$), the spacetime is singular, as
shown by Eq.(\ref{4.29a}). Thus, in the present case the spacetime is restricted to the region $|x| <  \sqrt{r_{g}}$, $ - \infty < t < \infty$ in the ($t, x$)-plane, with the two
spacetime singularities located at $x = \pm \sqrt{r_{g}}$ as its boundaries.
The global structure of the spacetime and  the corresponding Penrose diagram are  shown in Fig. \ref{fig6}. 

The change of variables (\ref{xrrelation}) can be understood by considering the one-form
$$e^\nu dr = \frac{\sqrt{r}dr}{\sqrt{D(r) (r_g -r)}}.$$
Even though the denominator of the right-hand side vanishes at $r = r_g$, we can turn $e^\nu dr$ into a nonsingular one-form by introducing a Riemann surface. Indeed, if we promote $r$ to a complex variable and define the genus $1$ Riemann surface $\Sigma$ as the two-sheeted cover of the complex $r$-plane obtained by introducing two branch cuts along the intervals $[0, r_g]$ and $[r_1, r_2]$, where $r_1$ and $r_2$ are the two (possibly complex) zeros of $D(r)$, $e^\nu dr$ is a holomorphic one-form on $\Sigma$. 
Letting $(0,r_g]_1$ and $(0, r_g]_2$ denote the covers of the interval $(0, r_g]$ in the first and second sheets of $\Sigma$, respectively, the spacetime consists of points $(r, \theta, \phi, t)$ with $r \in (0, r_g]_1 \cup (0, r_g]_2$.
The variable $x = - \sqrt{r_{g} - r}$ introduced in (\ref{xrrelation}) is analytic near the branch point at $r = r_g$ and $r \in (0, r_g]_1 \cup (0, r_g]_2$ corresponds to $x \in (-\sqrt{r_g}, \sqrt{r_g})$. We can fix the definition of $x$ by choosing the branch of the square root so that, say, $x \geq 0$ for $r \in (0, r_g]_1$. Thus, in terms of the variable $x$, the spacetime manifold can be covered by a single global chart (no double cover is necessary) and the metric $ds^2$, which involves the square of the differential $e^{\nu}dr$, is manifestly nonsingular at $r = r_g$. 
In particular, the metric of the extended spacetime is analytic, which ensures that the extension is unique.

The Hamiltonian constraint is
\begin{equation}\label{hamiltonianconstraintrg}  
  \int_0^{r_g} \mathcal{L}_V e^\nu r^2 dr = 0.
\end{equation}
Indeed, the Hamiltonian constraint (\ref{eq1}) should be interpreted as
\bq
  \int  \mathcal{L}_V \text{Vol}_g = 0,
\eq
where $\text{Vol}_g$ is the volume form induced by the metric $g_{ij}$ and the integration extends over a spatial slice of the spacetime. Using the variables $(r,\theta, \phi)$, we have
$$\text{Vol}_g = e^\nu r^2 \sin\theta dr d\theta d\phi,$$
and the integration extends over $\theta \in [0, \pi]$, $\phi \in [0, 2\pi]$, and $r \in [0, r_g]_1 \cup [0, r_g]_2$.
By symmetry, the contributions from the sets where $r \in [0, r_g]_1$ and $r \in [0, r_g]_2$ are equal. Since each contribution is proportional to the left-hand side of (\ref{hamiltonianconstraintrg}), the constraint reduces to (\ref{hamiltonianconstraintrg}). 

In view of (\ref{LVintegrand}), the constraint (\ref{hamiltonianconstraintrg}) becomes
\begin{equation}\label{Hamconstraintalphabeta}
\int_0^{r_g}
\frac{\alpha + \beta r + \gamma r^3 + \delta r^9}{36 r^7 \zeta^4 \sqrt{F(r)}} \sqrt{r} dr = 0.
\end{equation}
Denoting the integrand in (\ref{Hamconstraintalphabeta}) by $I(r)$, we see that $|I(r)|$ is bounded by a constant times $1/\sqrt{r_g -r}$ as $r \rightarrow r_g$. Thus, the integral converges near $r_g$. On the other hand, as $r \rightarrow 0$,
$$I(r) = \frac{\alpha}{36r^{\frac{13}{2}}\zeta^4\sqrt{C}} + O\left(\frac{1}{r^{\frac{11}{2}}}\right),$$
so that (\ref{Hamconstraintalphabeta}) can only be satisfied if $\alpha = 0$. Using similar arguments, we infer that the coefficients $\beta, \gamma$, $\delta$ must also vanish, i.e., 
$$\alpha = \beta = \gamma = \delta = 0.$$
Solving these equations, we conclude that the Hamiltonian constraint is satisfied if and only if the $g_j$'s satisfy the following four conditions:
\begin{eqnarray}\label{g6845conditions}
&&g_4 = \frac{\zeta ^4 (\Lambda_g-\Lambda )-2 g_2
   \zeta ^2 \Lambda_g^2}{4 \Lambda_g^3}, 
	\\ \nonumber 
&& g_5 =  -\frac{g_3 \zeta ^2}{2 \Lambda_g}, \qquad g_6 = 0, \qquad g_8 = 0.
\end{eqnarray}

Using the conditions (\ref{g6845conditions}) in the expression (\ref{Frrgeneral}) for $F_{rr}$, we find that Eqs. (\ref{4.3a}) and (\ref{4.3b}) have the solution
\begin{equation}\label{Aexpression}
A(r) = - \frac{\sqrt{F(r)}}{2\sqrt{r}} 
\int_{r_0}^r \frac{F_{rr}(r')(r')^{3/2} dr' }{\sqrt{F(r')}},
\end{equation}
where
\begin{eqnarray} \nonumber
 \label{Frrexpression}
 F_{rr} =&&\; 
-\frac{1} {36 r^8 \zeta ^2 \Lambda_g F(r)} 
 \Bigg\{-297 C^3 g_3   \nb\\
 & & -324C^2r g_3 + 126C^2 r^3\Lambda_g g_3
   	 \nonumber\\
&&   +12 C r^6 \Big[2\Lambda_g^2 (3 g_2+g_3)+\zeta ^2 (9 \Lambda -6\Lambda_g)\Big]
	\nonumber\\
&&  +8 r^9 \Lambda_g \Big[2\Lambda_g^2 (3 g_2+g_3)+\zeta ^2 (9 \Lambda -6 \Lambda_g)\Big]\Bigg\},\nb\\
 \end{eqnarray}  
 and $r_0 \in (0, r_g)$ is a constant.
The integrand in (\ref{Aexpression}) is smooth for $0 < r < r_g$. Thus, $A(r)$ is a smooth function of $r \in (0, r_g)$. Unless   $g_3 = 0$, the integral diverges as $r \rightarrow 0$, so that $A(r)$ has a singularity at $r = 0$. As $r \rightarrow r_g$, the integrand is bounded by $\text{const}\times(r_g - r)^{-3/2}$. This implies that $A(r)$ is bounded as $r \rightarrow r_g$. In fact, viewed as a function on the Riemann surface $\Sigma$, $A(r)$ is analytic near $r = r_g$. This follows since the integrand in (\ref{Aexpression}) is a meromorphic one-form with a pole of at most second order at $r = r_g$. Thus, the integral has a pole of at most order one at $r_g$, which is cancelled by the simple zero of the prefactor $\sqrt{F(r)} = \sqrt{D(r)(r_g - r)}$.
In conclusion, the gauge field $A$ given by (\ref{Aexpression}) is a smooth function everywhere on the extended spacetime away from the singularity at $r = 0$.

 \subsubsection{$C > 0,\; \Lambda_{g} < 0$}
In this case, $F(r) > 0$ for $r > 0$ and the metric coefficient $g_{rr}$ is positive and non-singular except at the point $r = 0$, at which a naked spacetime singularity appears. The  corresponding Penrose diagram is given by Fig. \ref{fig1} with $ r \in (0, \infty)$. The Hamiltonian constraint (\ref{4.3c}) requires that
\begin{equation}\label{constraintcase2}  
  \int_0^\infty \mathcal{L}_V e^{\nu} r^2 dr = 0.
\end{equation}
As in the previous subsection, this constraint is equivalent to the conditions given in (\ref{g6845conditions}). 

The function $A(r)$ is again given by the formulas (\ref{Aexpression})-(\ref{Frrexpression}) and is a smooth function of $r \in (0, \infty)$. As $r \to \infty$, the absolute value of the integrand is bounded by $\text{constant} \times r^{-2}$. Thus, choosing $r_0 = \infty$ in (\ref{Aexpression}), we find that $A(r)$ is bounded as $r \to \infty$.
Unless  $g_3 = 0$, the integral diverges as $r \to 0$, so that $A(r)$ has a singularity at $r = 0$.

\subsubsection{$C < 0,\; \Lambda_{g} > 0$}
\begin{figure}[tbp]
\centering
\includegraphics[width=8cm]{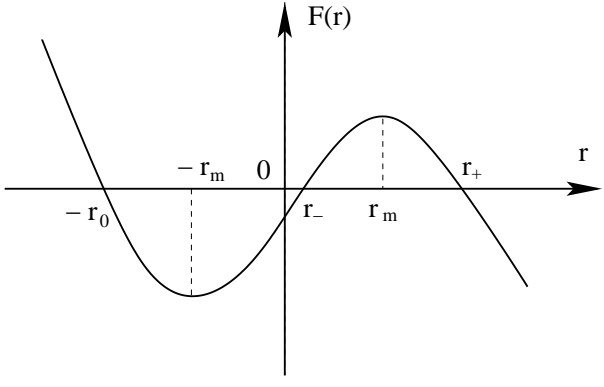}
\caption{ The function $F(r) = re^{-2\nu}$ for $C < 0$ and $\Lambda_{g} > 0$, where $r _{m} \equiv 1/\sqrt{\Lambda_{g}}$. $F(r) = 0$ has two positive roots $r_{\pm}$ only for $\Lambda_{g} < {4}/{(9C^{2})}$. When 
$\Lambda_{g} \ge {4}/{(9C^{2})}$,  $F(r)$ is always non-positive for any $r > 0$. }
\label{fig7}
\end{figure}

\begin{figure}[tbp]
\centering
\includegraphics[width=8cm]{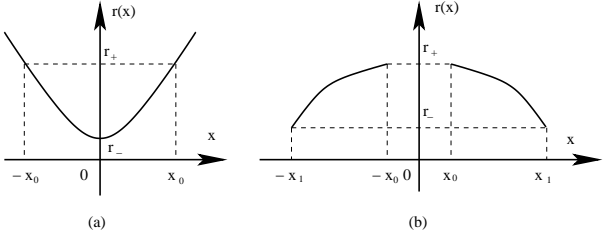}
\caption{ (a) The function $r$ vs $x$ given by Eq.(\ref{4.38a}), where $x_{0} \equiv \sqrt{r_{+}  - r_{-}}$. (b) The function $r$ vs $x$ given by Eq.(\ref{4.38b}), where $x_{1} \equiv  x_{0} + \sqrt{x_{0}}$.}
\label{fig8}
\end{figure}

In this case, if $\Lambda_{g} > 4/(9C^{2})$, $e^{2\nu} = r/F(r)$ is strictly negative  for all $r > 0$, so that, in addition to $t$, the coordinate $r$ is also timelike. The physics of such a spacetime
 is unclear,  if there is any. Therefore, in the following we consider only the case
\bq
\lb{4.36}
0 < \Lambda_{g} < \frac{4}{9C^{2}}.
\eq
 Then, we find that $F(r)$ is positive only for $0 < r_{-} < r < r_{+}$, where $r_{\pm}(\Lambda_{g}, C)$ are the two positive roots of   $F(r) = 0$, as shown in Fig. \ref{fig7}. We write $e^{2\nu}$ as
 \bq
 \lb{4.37}
 e^{2\nu} = \frac{r}{(r + r_{0})(r - r_{-})(r_{+} - r)},
 \eq
 where $r_{0}(\Lambda_{g}, C) > 0$. To extend the solution beyond $r = r_{\pm}$, we shall first consider the extension beyond $r = r_{-}$. Such an extension can be obtained via
 \bq
 \lb{4.38a}
 x = \int{\frac{dr}{2\sqrt{r - r_{-}}}} = \sqrt{r - r_{-}},
 \eq
 or inversely, $r = x^{2} + r_{-}$. Since $r < r_{+}$, we find that $ - x_{0} < x < x_{0}$ with  $x_{0} \equiv \sqrt{r_{+} - r_{-}}$. It can be seen that the coordinate singularity at $r = r_{-}$
 disappears, and the extended region is given by $ |x| < x_{0}$, as shown by Fig. \ref{fig8} (a).
 
\begin{figure}[tbp]
\centering
\includegraphics[width=8cm]{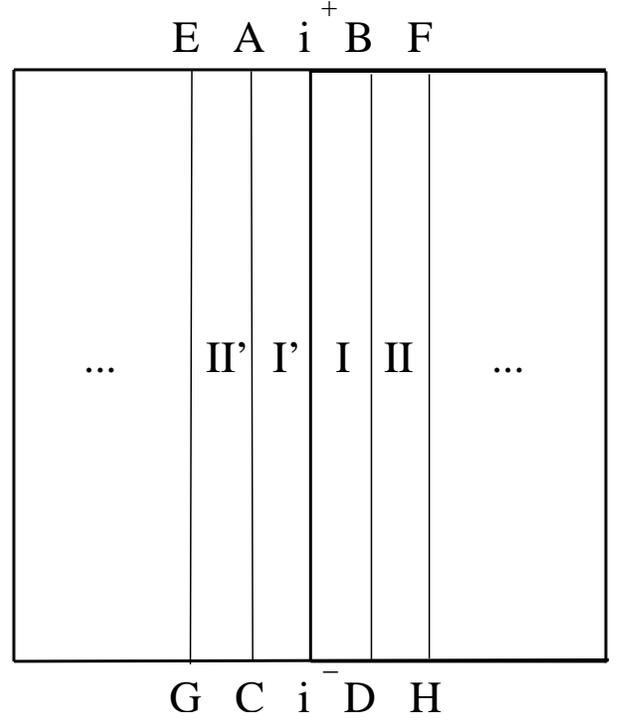}
\caption{ The global structure of the spacetime for $C < 0,\; \Lambda_{g} > 0$ and $\Lambda_{g} < 4/(9C^{2})$.  The vertical line $i^{+}i^{-}$ is the one where $r = r_{-}$, and the ones
$AC$ and $BD$ represent the lines where $r =r_{+}$, while on the lines $EG$ and $FH$ we have $r =r_{-}$. The spacetime   repeats itself  infinitely many times
in both directions of the $x$-axis. }
\label{fig9a}
\end{figure}

\begin{figure}[tbp]
\centering
\includegraphics[width=8cm]{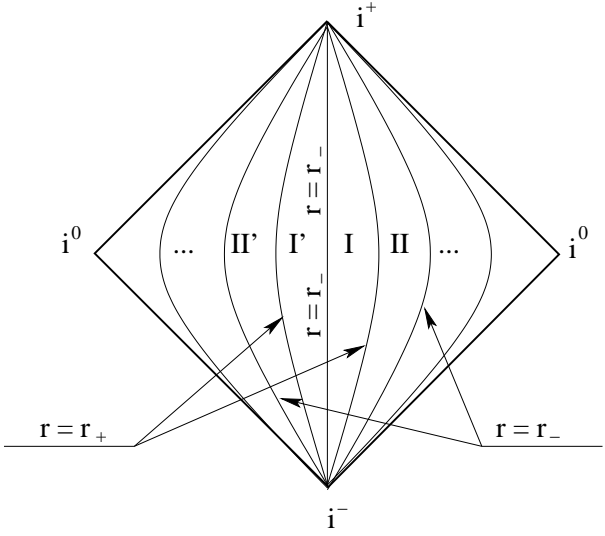}
\caption{ The Penrose diagram for $C < 0,\; \Lambda_{g} > 0$ and $\Lambda_{g} < 4/(9C^{2})$.  }
\label{fig9}
\end{figure}

To extend the solution beyond $r_{+}$, we introduce $x$ via the relation
 \bq
 \lb{4.38b}
 r = r_{+} - (x \mp x_{0})^{2}, 
 \eq
 where the ``$-$" sign applies when $x > x_0$ and the ``$+$" sign applies when $x < - x_{0}$. Fig. \ref{fig8} (b) shows the graph of $r$ as a function of $x$. From Fig. \ref{fig8}   we can see that the extension along  both the positive and the negative 
 directions of $x$ need to continue in order to have a    maximal spacetime. This can be done by repeating the above process infinitely many times, so finally   the whole ($t, x$)-plane 
 is covered by an infinite number of finite strips, in each of which we have $ r_{-} \le r \le r_{+}$. The global structure is that of
 Fig. \ref{fig9a} and the corresponding Penrose diagram is given by Fig. \ref{fig9}. Thus, in this case we have $r \in[r_{-}, r_{+}]$.

The Hamiltonian constraint (\ref{4.3c}) requires that
\begin{equation}\label{constraintcase3}
\int_{r_-}^{r_+}
\frac{\alpha + \beta r + \gamma r^3 + \delta r^9}{36 r^7 \zeta^4 \sqrt{F(r)}} \sqrt{r} dr = 0.
\end{equation}
Geometrically, this condition can be understood by introducing a Riemann surface $\Sigma$ as a double cover of the complex $r$-plane with two branch cuts along $[r_-, r_+]$ and $[-r_0, 0]$. The integrand in (\ref{constraintcase3}) is a one-form $\omega$ on $\Sigma$ which is holomorphic in a neighborhood of the closed curve $a_1 \equiv [r_-, r_+]_1 \cup [r_+, r_-]_2$. Topologically, the elliptic curve $\Sigma$ is a torus, $a_1$ is a nontrivial cycle, and the condition (\ref{constraintcase3}) states that the integral of $\omega$ along the cycle $a_1$ vanishes. 
This imposes a constraint on the coefficients $\alpha, \beta, \gamma, \delta$, which translates into a condition on the $g_j$'s involving elliptic integrals. Assuming this condition to hold, the function $A(r)$ is given by (\ref{Aexpression}) with $r_0 \in (r_-, r_+)$ and $F_{rr}$ as in (\ref{Frrgeneral}).

\subsubsection{$C < 0,\; \Lambda_{g} < 0$} 
\begin{figure}[tbp]
\centering
\includegraphics[width=8cm]{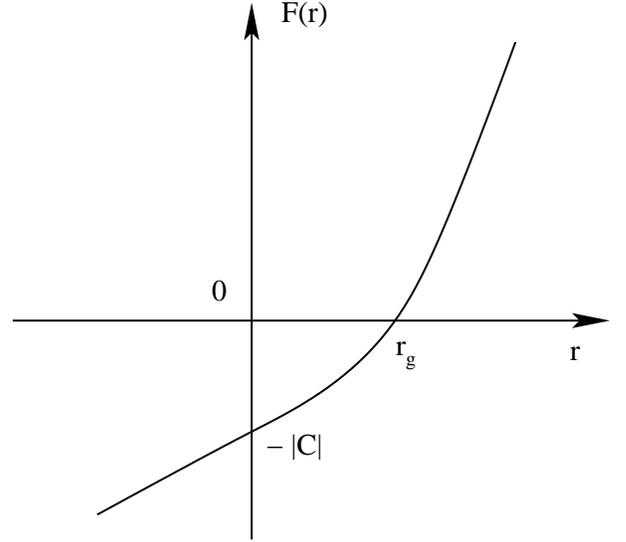}
\caption{ The function $F(r) = re^{-2\nu}$ for $C < 0$ and $\Lambda_{g} < 0$, where $r_{g}$ is the only positive root of $F(r) = 0$. }
\label{fig10}
\end{figure}

In this case,  the function $F(r) =  re^{-2\nu}$ is positive only for $r > r_{g}$, as shown in Fig. \ref{fig10}. Thus,   $e^{2\nu}$ can be written in the form,
\bq
\lb{4.40}
e^{2\nu} = \frac{r}{D(r) (r - r_{g})},
\eq
where $D(r) > 0$ for $r > 0$. The extension can be carried out by introducing a new coordinate $x$
via the relation,
\bq
\lb{4.41}
r = x^{2} + r_{g}.
\eq
In terms of $x$ the coordinate singularity at $r = r_{g}$ disappears, and the extended spacetime is given by $ - \infty < t, x < \infty$ in the ($t, x$)-plane. Its global structure
is given by Fig. \ref{fig3a}, while 
the  corresponding Penrose diagram  is given by Fig. \ref{fig3}. Thus, in this case the range of $r$ is  $r \in [r_{g}, \infty)$. 

The Hamiltonian constraint (\ref{4.3c}) requires that
$$\int_{r_g}^{\infty}
\frac{\alpha + \beta r + \gamma r^3 + \delta r^9}{36 r^7 \zeta ^4 \sqrt{F(r)}} \sqrt{r} dr = 0.$$
The behavior of the integrand as $r \to \infty$ implies that
$$\alpha = \beta = \gamma = \delta = 0,$$
so that the constraint reduces to (\ref{g6845conditions}) and the function $A(r)$ is given by (\ref{Aexpression})-(\ref{Frrexpression}),
which  is  not singular  everywhere in the extended spacetime.

\section{Vacuum Solutions with $N^{r} \not= 0$}
\renewcommand{\theequation}{5.\arabic{equation}} \setcounter{equation}{0}

When $N^{r} \not= 0$, the vacuum solutions are given by  \cite{GSW},  
\bq
\lb{5.0}
ds^{2} = - dt^{2} + e^{2\nu}\left(dr + e^{\mu-\nu} dt\right)^{2} + r^{2}d^{2}\Omega,
\eq
  with
 \bqn
 \lb{5.1}
 \mu &=& \frac{1}{2}\ln\Bigg(\frac{2m}{r} + \frac{1}{3}\Lambda r^2 - 2A(r) + \frac{2}{r}\int^{r}{A(r') dr'}\Bigg),\nb\\
  \nu &=&   \varphi = \Lambda_{g} = 0,
 \eqn
 where the gauge field $A$ must satisfy   the  Hamiltonian constraint,
\bq
\lb{5.2}
\int_{0}^{\infty}{r A'(r) dr} = 0.
\eq
Otherwise, it is free. However, as shown in \cite{GSW}, the solar system tests seem uniquely to choose the Schwarzschild 
solution $A = 0$. Therefore, in the following we shall consider only this case, 
 \bqn
 \lb{5.1a}
 \mu &=& \frac{1}{2}\ln\Bigg(\frac{2m}{r} + \frac{1}{3}\Lambda r^2\Bigg),\nb\\
  \nu &=&   \varphi = \Lambda_{g} = A = 0.
 \eqn
 It should be noted that if $(N, \nu, N^{r})$ is a solution of the vacuum equations, so is $(N, \nu, - N^{r})$. The latter can be
 easily obtained by the replacement  $t \rightarrow - t$. With such changes,  we have $K_{ij} \rightarrow - K_{ij}$ (in the static case).
Clearly, these   do not affect the singularity behavior. 
 We then obtain \cite{CW,GPW} \footnote{There is a typo in the expression of K given by Eq.(3.2) in \cite{CW}. Although it propagates 
to other places,  this does not affect our main conclusions, as $K$ and $K_{ij}K^{ij}$ have similar singularity behavior. },
\bqn
\lb{5.1b}
R_{ij} &=& 0,\nb\\
K &=& \epsilon_{1} \sqrt{\frac{3}{r^{3}\left(6m + \Lambda r^{3}\right)}} \; \left(3m + \Lambda r^{3}\right),\nb\\
K_{ij}K^{ij} &=&  \frac{27m^{2} + 6m\Lambda r^{3} + \Lambda^{2}r^{6}}{r^{3}\left(6m + \Lambda r^{3}\right)},
\eqn
where $\epsilon_{1} (= \pm 1)$ originates from  the expression $N^{r} = \epsilon_{1} e^{\mu}$, obtained by the 
replacement  $t \rightarrow - t$, as mentioned above.
To further study  the above solutions,  let us  consider the cases
(1) $\; m = 0, \Lambda \not=0$;  (2) $\; m \not= 0, \Lambda =0$; and (3) $\; m \not= 0, \Lambda \not=0$ separately. We shall
assume that $m \ge 0$, while $\Lambda$ can take any values. 

\subsection{$m = 0, \;\;\; \Lambda \not=0$}

In this case, only $\Lambda > 0$ is allowed \cite{GSW}, as can be seen from
Eq.(\ref{5.1}). That implies that the anti-de Sitter spacetime cannot be written in the static form of Eq.(\ref{3.1b})
with the  projectability condition. Then, we have $ N^{2} = f = 1, \; N^{r} = \epsilon_{1} {r}/{\ell}$, or
\bq
\lb{5.3}
ds^2 = - dt^{2} + \left(dr + \epsilon_{1} \frac{r}{\ell} dt\right)^{2} + r^{2}d^{2}\Omega,
\eq
where $\ell \equiv \sqrt{3/|\Lambda|}$.  
Without loss of generality, we shall consider only the case $\epsilon_{1} = -1$, as the case   
 $\epsilon_{1} = 1$ can be simply obtained from the one $\epsilon_{1} = -1$ by inverting the time coordinate. 
In terms of $N, N^{i}, g_{ij}$ or their inverses, $N_{i}, g^{ij}$, the metric is non-singular, except for the trivial  $r = 0$ and
$\theta = 0, \pi$. In addition,   from Eq.(\ref{5.1b}) we also find that 
\bq
\lb{5.5}
K = -\sqrt{2\Lambda},\; \;\; K_{ij}K^{ij} = \Lambda, \; (m = 0).
\eq
On the other hand,  in terms of the 4-dimensional metric, $g_{\mu\nu}$ and $g^{\mu\nu}$, it is  not singular either, as
one can see from the expressions,
\bqn
\lb{5.3a}
\left({}^{(4)}g_{\mu\nu}\right) &=&  \left(\matrix{ - \frac{\ell^{2} - r^{2}}{\ell^{2}}, &- \frac{r}{\ell}\delta^{r}_{i}\cr
-\frac{r}{\ell}\delta^{r}_{i}, & g_{ij}}\right),\nb\\
\left({}^{(4)}g^{\mu\nu}\right) &=& \left(\matrix{ - 1, & -\frac{r}{\ell}\delta_{r}^{i}\cr
-\frac{r}{\ell}\delta_{r}^{i}, & g^{ij} - \frac{r^{2}}{\ell^{2}}\delta_{r}^{i}\delta_{r}^{j}\cr}\right),
\eqn
although the nature of the radial coordinate does change,
\bq
\lb{5.3b}
g^{\mu\nu}r_{,\mu} r_{,\nu} = 1 - \frac{r^{2}}{\ell^{2}} = \cases{{\mbox{ timelike}}, & $ r > \ell$, \cr
{\mbox{ null}}, & $ r = \ell$, \cr
{\mbox{ spacelike}}, & $ r < \ell$. \cr}
\eq
To study  the   solution further in the HL theory, we consider two different regimes,
$E \ll M_{*}$ and  $E \gg M_{*}$,  where $M_{*} = {\mbox{min.}}\left\{M_{A}, \; M_{B}, ... \right\}$ and $M_{n}$'s are the energy scales
appearing in the dispersion relation (\ref{Poly}).

\subsubsection{$E \ll M_{*}$}

When the energy $E$ of the test particle is much less than $M_{*}$, from Eq.(\ref{Poly}) one can see that $F(\zeta) \simeq \zeta$. 
This corresponds to the relativistic case (n = 1), studied in Sec. III.A.2. Then, for the ingoing test particles ($\epsilon = -1$), we have
\bq
\lb{5.4}
H = N\sqrt{f} + N^{r} = \frac{\ell - r}{\ell} .
\eq
Thus, the hypersurface $r = \ell$ is indeed a horizon. In fact, it represents a cosmological horizon, as first found in GR \cite{GH}. 

However, because of the restricted diffeomorphisms (\ref{1.2}), it is very interesting to see the global structure of the 
de Sitter spacetime in the HL theory. To this purpose, let us consider the coordinate transformations,
\bq
\lb{5.6}
t' = \ell e^{-t/\ell},\;\;\;
r' = r e^{-t/\ell},
\eq
in terms of which the corresponding metric takes the form,
\bqn
\lb{5.7}
ds^{2}&=& - d{t}^{2} + e^{2{t}/\ell}\left(dr'^{2} + r'^{2}d^{2}\Omega\right).\nb\\
&=&  \left(\frac{\ell}{t'}\right)^{2}\left(- dt'^{2} + dr'^{2} + r'^{2}d^{2}\Omega\right).
\eqn
From Eq.(\ref{5.6}) we can see that the whole $(t, r)$-plane, $ -\infty < t < \infty,\; r \ge 0$, is mapped to the region $t', \; r ' \ge 0$.  However, the metric
now becomes singular at $t' = 0, \infty$ (or $t = \pm \infty$). To see the nature of these singularities, one may recall the 5-dimensional embedding of
the de Sitter spacetime in GR \cite{HE73}, from which we find that in terms of the 5-dimensional coordinates $v$ and $w$, $t'$ is given by
 $t' = \ell^{2}/(v + w)$. Therefore, $t'  \ge 0$ corresponds to $v + w \ge 0$.  Thus,  the region $t', \; r ' \ge 0$  only represents the half hyperboloid 
 $v + w \ge 0$, as shown by Fig. 16 (ii) in \cite{HE73}.  In particular,  $t' = 0$ represents the boundary of the
 spacelike infinity, so extension beyond this surface may not be needed.  Although
 the extension given in \cite{HE73} in terms of the  static Einstein universe coordinates
 $(\bar{t}, \bar\chi, \bar{\theta},\bar{\phi})$  is forbidden here by the restricted diffeomorphisms (\ref{1.2}),
  as that extension requires, 
 $$
 t = \ell \ln\left[\cosh\left(\frac{\bar{t}}{\ell}\right)\cos(\bar\chi) +  \sinh\left(\frac{\bar{t}}{\ell}\right)\right],
 $$
  the extension across $t' = \infty$ (or $v + w = 0^{+}$)  seems necessary. 

\begin{figure}[tbp]
\centering
\includegraphics[width=8cm]{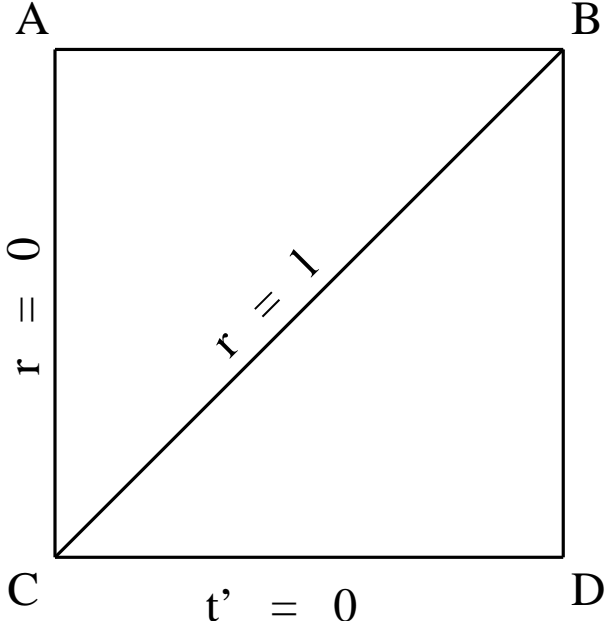}
\caption{ The global structure of the de Sitter solution $N^{2} = f = 1, \; N^{r} = - \sqrt{r/\ell}$ in the HL theory with the restricted diffeomorphisms (\ref{1.2}) for the 
region $t' \ge 0$. The horizontal line  $AB$ corresponds to $t' = \infty$ (or $t = -\infty$), while the vertical line $BD$ to $r' = \infty$ (or $r = \infty$).}
\label{fig11}
\end{figure}

 Another way to see the need of an extension beyond $t' = 0$ is that the metric (\ref{5.7}) is well-defined for $t' < 0$. So, one may  simply  take   $ - \infty < t' < \infty$. 
 But, this cannot be considered as an extension, as   the metric (\ref{5.7}) is singular at $t' = 0$, and the two regions $t' > 0$ and $t' < 0$ are not smoothly connected
 in the $t', r'$-coordinates.  In this sense, a proper extension is still needed. However, due to the  restricted diffeomorphisms (\ref{1.2}), it is not clear if such extensions
 exist or not.  Fig. \ref{fig11} shows the global structure of the region $t' \ge 0$, which is quite different from its corresponding Penrose diagram \cite{GH}.

\subsubsection{$E \gg M_{*}$}

When the energy $E$ of the test particle is greater than  $M_{*}$,  from Eq.(\ref{Poly}) one can see that high order momentum terms become important, and
$F(\zeta) \simeq \zeta^{n}, \; (n \ge 2)$. For the sake of simplicity,  we consider the case with $n = 2$ only. Then, from Eqs.(\ref{eqq16}) and 
(\ref{4.19}) we find that 
\bqn
\lb{5.8}
X &=& \frac{2\ell E}{\sqrt{r^{2} + 4 \ell^{2}E} + r},\nb\\
H &=& \frac{r}{\ell} -  \frac{4\ell E}{\sqrt{r^{2} + 4 \ell^{2}E} + r}.
\eqn
Thus, $H(r, E) = 0$ has only one real root,
\bq
\lb{5.9}
r_{H} = \left(\frac{4\ell^{2}E}{3}\right)^{1/2},
\eq
at which we find that
\bq
\lb{5.10}
H(r_{H}, E) =  \frac{12\ell E}{4 \ell^{2}E + 3r^{2}_{H}} > 0. 
\eq
Eqs.(\ref{CT}) and (\ref{delta}) then tell us that the surface $r = r_{H}$ is a horizon for a test particle with energy $E$. It should be noted that, in contrast to the Schwarzschild case studied
in Sec. III.B.2 [cf. Eq.(\ref{root})],  $r_{H}$ now is proportional to $E$, that is, the higher the energy of the test particle, the lager  the radius of the horizon. To understand
this, let us consider the acceleration of a test particle with its four-velocity $u_{\lambda} = - \delta^{t}_{r}$,  located on a surface $r$. Then, we find that
\bq
\lb{acc}
a_{\mu} \equiv u_{\mu;\lambda} u^{\lambda} = \cases{- \frac{m}{r^{2}}\delta^{r}_{\mu}, & Schwarzschild,\cr
\frac{r}{\ell^{2}}\delta^{r}_{\mu}, & de Sitter.\cr}
\eq
That is, for the Schwarzschild solution, the test particle feels an attractive force, while for the de Sitter solution, it feels a repulsive one. Because of this difference,  in the 
de Sitter spacetime $r_{H}$ is proportional to $E$, in contrast to the Schwarzschild one, where it is inversely proportional to $E$, as shown explicitly in Eq.(\ref{root}).

\subsection{$m > 0, \;\;\; \Lambda =0$}

When $\Lambda = 0$ and $m > 0$, it is the Schwarzschild solution studied in  Sec. III.B.2 and Sec.III.C in detail. In particular, in the IR, the surface $r = 2m$ represents a
horizon, while for high energy particles, the radius of the horizon is energy-dependent, as explicitly given by Eq.(\ref{root}) for $n = 2$.
So, we shall not repeat these studies, but simply note that
now the solution takes the form,
\bq
\lb{5.11}
ds^{2} = - dt^{2} + \left(dr - \sqrt{\frac{2m}{r}}dt\right)^{2} + r^{2}d\Omega^{2},
\eq
which is singular only at $r = 0$, as can be seen from Eq.(\ref{5.1b}). So, it already represents a maximal spacetime
in the HL theory. 

It is interesting to note that the above metric covers only  half  of the maximally extended spacetime given in GR. This can be seen easily by introducing 
the coordinate $\tau$ \cite{GPW},
\bqn
\lb{5.12}
\tau & \equiv& t - \int{\frac{\sqrt{2mr}}{r - 2m} dr}\nb\\
&=& t - 2\sqrt{2mr}  - 2m \ln\left(\frac{r - 2m}{\left(\sqrt{r} + \sqrt{2m}\right)^{2}}\right),~~~
\eqn
in terms of which, the solution takes the standard Schwarzschild form,
$ds^{2} = - f(r)d\tau^{2} + f^{-1}(r)dr^{2} + r^{2}d\Omega^{2}$ with $f(r) = 1 - 2m/r$. Of course, the above transformations are forbidden by Eq.(\ref{1.2}).

 \subsection{$m > 0, \;\;\; \Lambda \not= 0$}
 
 In this case, it is convenient to further distinguish the two subcases $\Lambda > 0$ and $\Lambda < 0$.
 
 \subsubsection{$\Lambda > 0$}
 
 In this case, the metric takes the form, 
 \bq
 \lb{5.13a}
 ds^{2} = -dt^{2} + \left(dr - \sqrt{\frac{2m}{r} + \frac{r^{2}}{\ell^{2}}} \;dt\right)^{2} + r^{2}d\Omega^{2}.
 \eq
 
 When $E \ll M_{*}$, as in the last case the dispersion relation becomes relativistic, and $F(\zeta) \simeq \zeta$, for which we have $n = 1$. Then, we find that
 \bqn
 \lb{5.13}
 H(r)  &=& 1 + N^{r} = 1 - \sqrt{\frac{2m}{r} + \frac{r^{2}}{\ell^{2}}} \nb\\
 &=& \frac{F(r)}{\ell^{2}\left(1 + \sqrt{\frac{2m}{r} + \frac{r^{2}}{\ell^{2}}}\right)},
 \eqn
 but now $F(r) \equiv - \left(r^{3} - \ell^{2}r + 2m \ell^{2}\right)$. Clearly, $F(r)$ has one maximum and one minimum, respectively, at $r = \pm r_{m}$, where
 $r_{m} =  \ell/\sqrt{3}$ and $F(r_{m}) = -2\ell^{2}(m - 1/(3\sqrt{\Lambda})$, as shown in Fig. \ref{fig7}.
 Thus, when $ m^{2} > 1/(9\Lambda^{2})$, $H(r) = 0$ has no real positive root, and a horizon does not exist even in the IR. Therefore, the singularity at $r = 0$ is
 naked. When $ m^{2} < 1/(9\Lambda^{2})$, $H(r) = 0$ has two real and positive roots,  $r_{\pm},\; (r_{+} > r_{-})$, where $r = r_{+}$ is often referred to as
 the cosmological horizon and $r = r_{-}$ the black hole event horizon \cite{GH}. When $m^{2} = 1/(9\Lambda^{2})$, the two horizons coincide. In GR, the 
 corresponding Penrose diagrams were given in \cite{GH}. However, as argued above, in the HL theory these diagrams are not allowed, as they are obtained
 by coordinate  transformations that violate the restricted diffeomorphisms (\ref{1.2}). Nevertheless, since the metric is not singular in the current form, it already 
 represents a maximal  spacetime. 
 
 When $E \gg M_{*}$, the high momentum terms dominate, and for $n = 2$, we find that
  \bqn
 \lb{5.14}
 X(r)  &=&  \frac{2E}{\sqrt{\frac{2m}{r} + \frac{r^{2}}{\ell^{2}}+ 4E} + \sqrt{\frac{2m}{r} + \frac{r^{2}}{\ell^{2}}}},  \nb\\
H(r) &=& \sqrt{\frac{2m}{r} + \frac{r^{2}}{\ell^{2}}} -  2X = \frac{F(r)}{\Delta(r)},
 \eqn
 where $\Delta(r) > 0$ for any $r \in (0, \infty)$, and $F(r) \equiv r^{3} - 4E\ell^{2}r/3 + 2m \ell^{2}$. 
 It can be shown that when $ m^{2} > 8\ell E^{3/2}/27$,  $H(r) = 0$ has no real and positive roots. Thus, in this case there are no  horizons, and the singularity at
 $r = 0$ must be naked.  When $ m^{2} < 8\ell E^{3/2}/27$,  $H(r) = 0$ has two real and positive roots, say, $r_{1, 2} \; (r_{2} > r_{1})$, but now $r_{1, 2} = r_{1,2}(E, m, \ell)$.
 Thus,  in this case there also exists two horizons, but each of them depends on $E$.  When $ m^{2} = 8\ell E^{3/2}/27$, we have $r_{1} = r_{2}$, and the two horizons 
 coincide.

 \subsubsection{$\Lambda < 0$}

 In this case, the metric takes the form, 
 \bq
 \lb{5.15}
 ds^{2} = -dt^{2} + \left(dr - \sqrt{\frac{2m}{r} - \frac{r^{2}}{\ell^{2}}} \;dt\right)^{2} + r^{2}d\Omega^{2},
 \eq
 where $\ell \equiv \sqrt{3/|\Lambda|}$. Then, from Eq.(\ref{5.1b}), it can be seen that the spacetime is singular at
 $r_{s} \equiv (2m\ell^{2})^{1/3}$ \cite{CW}. This is different from GR, in which the only singularity of the anti-de Sitter Schwarzschild solution is at $r = 0$. 
 
  When $E \ll M_{*}$, as in the last case the dispersion relation becomes relativistic. Then, we find that
 \bqn
 \lb{5.16}
 H(r)  &=& 1 + N^{r} = 1 - \sqrt{\frac{2m}{r} - \frac{r^{2}}{\ell^{2}}} \nb\\
 &=&  \frac{F(r)}{r\ell^{2}\left(1 + \sqrt{\frac{2m}{r} - \frac{r^{2}}{\ell^{2}}}\right)},
 \eqn
 but now with $F(r) \equiv r^{3} + \ell^{2}r - 2m \ell^{2}$, which is a monotonically increasing function, as shown by Fig. \ref{fig10}. 
 Thus, $H(r) = 0$  has one and only one  real and positive root $r_{H} = r_{H}(m, \ell)$. But, $r_{H}$ is 
 always less than $r_{s}$, i.e., $r_{H} < r_{s}$. Thus, the singularity 
 at $r= r_{s}$ is a naked singularity.
 
  When $E \gg M_{*}$, let us consider only the case $n = 2$. Then,  we find that
   \bqn
 \lb{5.17}
 X(r)  &=&  \frac{2E}{\sqrt{\frac{2m}{r} - \frac{r^{2}}{\ell^{2}}+ 4E} + \sqrt{\frac{2m}{r} - \frac{r^{2}}{\ell^{2}}}},  \nb\\
H(r) &=& \sqrt{\frac{2m}{r} - \frac{r^{2}}{\ell^{2}}} -  2X = \frac{F(r)}{\Delta(r)},
 \eqn
 where $\Delta(r) > 0$ for any $r \in (0, \infty)$, and $F(r) \equiv r^{3} + 4E\ell^{2}r/3 - 2m \ell^{2}$. 
 It can be shown that this $F(r)$ is also a monotonically increasing function, as shown by Fig. \ref{fig10}, and $F(r) = 0$ has only one real and positive root,
  $r_{H} = r_{H}(m, E, \ell)$. Again, since $H(r_{s}) = 1$ and $H(r_{H}) = 0$, we find that $r_{H}$ is 
 also always less than $r_{s}$,  although now  $r_{H}$ depends on $E$. Thus, the singularity 
 at $r= r_{s}$ is a naked singularity.

\section{Conclusions}
 \renewcommand{\theequation}{7.\arabic{equation}} \setcounter{equation}{0}

 In this paper, we have systematically studied black holes in the HL theory, using the kinematic method of test particles provided by  KK  in \cite{KKb}, in which a horizon 
 is defined as the surface at which massless test particles are infinitely redshifted. Because of the nonrelativistic  dispersion relations (\ref{1.4}), in Sec. III we have shown explicitly the 
 difference between   black holes defined in GR and the ones defined here. In particular, the radius of the horizon  usually depends on the energy of the test particles.

 When applying this definition to the spherically symmetric and static vacuum solutions found recently in \cite{HMT,AP,GSW},   in  Secs. IV and V we 
 have found  that   for test particles with sufficiently high energy, the radius of the horizon can be made arbitrarily small,
 although the singularities at the center can be seen in principle only by test particles with  infinitely high energy. 
 
 In Secs. IV and V, we  paid particular attention to the global structures of the static solutions. Because of the restricted diffeomorphisms (\ref{1.4}), they
 are dramatically different from the corresponding ones given in GR, even  the solutions are the same. In particular, the restricted diffeomorphisms (\ref{1.4}) do not allow us to draw Penrose
 diagrams, although one can create  something similar to them,  for example, see Figs. \ref{fig1a}, \ref{fig3a}, \ref{fig6}, \ref{fig9a}, \ref{fig11}. 
 But,  it must be noted that, since the speed of the test particles in the HL theory can 
 be infinitely large, the causality in this theory is also dramatically different from that of GR [cf. Fig.\ref{fig0}]. In particular, the light-cone structure in GR does not apply to the
 HL theory.  Among the static solutions, a very interesting case is the one given by Fig. \ref{fig3a}, which corresponds to an Einstein-Rosen bridge. In GR, this solution
is made of an exotic fluid as one can see from Eq.(\ref{4.17b}), which is clearly unphysical, and most likely unstable, too. However, in the HL theory, the solution is a vacuum one,
and it would be very interesting to see if this configuration is stable or not in the HMT setup.

 Finally, in Appendix B we have studied the slowly rotating solutions in the HMT setup \cite{HMT}, and found explicitly all such solutions, which are
  characterized by an arbitrary function $A_{0}(r)$. When   $A_{0} = 0$ they reduce    to the slowly rotating Kerr solution obtained in GR. 
  When the rotation is switched off, they reduce to the static solutions obtained in \cite{GSW}.

~\\{\bf Acknowledgements:}   JXL would like to thank  the Physics Department and CASPER at 
Baylor University for hospitalities during his visit there where part of this work was initiated and completed.
Part of this work was also done when two of the authors (JXL $\&$ AW)  attended
the advanced workshop ``Dark Energy and Fundamental Theory," Tunxi,  China, April 8 - 18, 2011, supported by the Special Fund for Theoretical Physics
 from the National Natural Science Foundation  of China (NNSFC) by the grant 10947203. 
 JL acknowledges support from the EPSRC, UK. JXL acknowledges support from the Chinese Academy of Sciences, a grant
from 973 Program with grant No: 2007CB815401 and a grant from the NNSFC
with Grant No : 10975129.  AW  is supported in part by DOE Grant, DE-FG02-10ER41692 and NNSFC grant, 11075141.

 \section*{Appendix A: The Functions $\left(F_{s}\right)_{ij}$}
 \renewcommand{\theequation}{A.\arabic{equation}} \setcounter{equation}{0}
 
 For the solution
 \bq
 \lb{A.1}
 \nu = - \frac{1}{2} \ln\left(1 + \frac{C}{r} -\frac{1}{3} \Lambda_g r^2\right),
 \eq
 the functions  $\left(F_{s}\right)_{ij}$  appearing in Eq.(\ref{eq3a}) are given by
 \bqn
 \lb{A.2}
\left(F_{0}\right)_{ij} &=& -\frac12g_{ij}\nb
	\\
&=& - \frac{1}{2}e^{2\nu}\delta^{r}_{i} \delta^{r}_{j} - \frac{1}{2}r^{2}\Omega_{ij},\nb\\
\left(F_{1}\right)_{ij} &=&-\frac12g_{ij}R+R_{ij} \nb
	\\
&=& \frac{e^{2\nu}}{3r^{3}}\left(3C - \Lambda_{g}r^{3}\right)\delta^{r}_{i} \delta^{r}_{j}
 - \frac{1}{6r}\left(3C +2 \Lambda_{g}r^{3}\right) \Omega_{ij},\nb\\
\left(F_{2}\right)_{ij} &=&-\frac12g_{ij}R^2+2RR_{ij}-2\nabla_{(i}\nabla_{j)}R +2g_{ij}\nabla^2R\nb\\
 &=& \frac{2\Lambda_{g}e^{2\nu}}{3r^{3}}\left(6C + \Lambda_{g}r^{3}\right)\delta^{r}_{i} \delta^{r}_{j}
   \nb\\
   & & ~~~~~~ - \frac{2\Lambda_{g}}{3r}\left(3C - \Lambda_{g}r^{3}\right) \Omega_{ij},\nb\\
\left(F_{3}\right)_{ij} &=&-\frac12g_{ij}R_{mn}R^{mn}+2R_{ik}R^k_j-2\nabla^k\nabla_{(i}R_{j)k}\nb\\
                     && +\nabla^2R_{ij}+g_{ij}\nabla_m\nabla_nR^{mn}\nb
	\\
&=& \frac{e^{2\nu}}{36r^{6}}\left(-9C^2  + 60 C \Lambda_{g}r^{3} + 8 \Lambda_{g}^{2} r^{6}\right)\delta^{r}_{i} \delta^{r}_{j}
   \nb\\
   & & ~~~ + \frac{1}{18r^{4}}\left(9C^2  -15 C \Lambda_{g}r^{3} + 4 \Lambda_{g}^{2} r^{6}\right) \Omega_{ij},\nb\\
\left(F_{4}\right)_{ij} &=&-\frac12g_{ij}R^3+3R^2R_{ij}-3\nabla_{(i}\nabla_{j)}R^2\nb\\
                    && +3g_{ij}\nabla^2R^2\nb
	\\                    
&=& \frac{4\Lambda_{g}^{2}e^{2\nu}}{r^{3}}\left(3C  + \Lambda_{g}r^{3}\right)\delta^{r}_{i} \delta^{r}_{j}
   \nb\\
   & & ~~~~~~ - \frac{2\Lambda_{g}^{2}}{r}\left(3C  -2 \Lambda_{g}r^{3}\right) \Omega_{ij},\nb\\
\left(F_{5}\right)_{ij} &=&-\frac12g_{ij}RR^{mn}R_{mn}+R_{ij}R^{mn}R_{mn}\nb\\
                    &&+ 2RR_{ki}R^k_j  -\nabla_{(i}\nabla_{j)}\left(R^{mn}R_{mn}\right)\nb\\
                    && - 2\nabla^n\nabla_{(i}RR_{j)n}  +g_{ij}\nabla^2\left(R^{mn}R_{mn}\right)\nb\\
                    &&  + \nabla^2\left(RR_{ij}\right)   +g_{ij}\nabla_m\nabla_n\left(RR^{mn}\right)\nb\\
&=& \frac{e^{2\nu}}{6r^{9}}\Big(-99 C^3+39 C^2 \Lambda_g r^3-108 C^2 r
    \nb\\
& & ~~~~~~~~ +28 C \Lambda_g^2 r^6+8
   \Lambda_g^3 r^9 \Big)\delta^{r}_{i} \delta^{r}_{j}
   \nb\\
   & & ~~ + \frac{1}{12r^{7}}\Big(693 C^3-156 C^2 \Lambda_g r^3+648 C^2 r \nb\\
& & ~~~~~~~~~~~~~~ -28 C \Lambda_g^2 r^6+16
   \Lambda_g^3 r^9\Big)\Omega_{ij},\nb\\
\left(F_{6}\right)_{ij} &=&-\frac12g_{ij}R^m_nR^n_pR^p_m+3R^{mn}R_{ni}R_{mj}\nb\\
                      && +\frac32\nabla^2\left(R_{in}R^n_j\right)   + \frac32g_{ij}\nabla_k\nabla_l\left(R^k_nR^{ln}\right)\nb\\
                      & & -3\nabla_k\nabla_{(i}\left(R_{j)n}R^{nk}\right)\nb\\
&=& \frac{e^{2\nu}}{36r^{9}}\Big(-675C^3  +270 C^{2}\Lambda_{g} r^{3} -729 C^{2} r \nb\\
& & ~~~~~~~~ + 72C \Lambda_{g}^{2} r^{6} 
       + 16 \Lambda^{3}_{g} r^{9}\Big)\delta^{r}_{i} \delta^{r}_{j}\nb\\
& & ~~ + \frac{1}{72r^{7}}\Big(4725C^3  - 1080 C^{2}\Lambda_{g} r^{3} + 4374 C^{2} r \nb\\
& & ~~~~~~~~~~~~~~ - 72C \Lambda_{g}^{2} r^{6} 
       + 32 \Lambda^{3}_{g} r^{9}\Big)\Omega_{ij},\nb\\
\left(F_{7}\right)_{ij} &=&\frac12g_{ij}(\nabla R)^2- \left(\nabla_iR\right)\left(\nabla_jR\right) +2R_{ij}\nabla^2R\nb\\
&& -2\nabla_{(i}\nabla_{j)}\nabla^2R+2g_{ij}\nabla^4R\nb\\
&=& 0,\nb\\ 
\left(F_{8}\right)_{ij} &=& -\frac12g_{ij}\left(\nabla_p R_{mn}\right)\left(\nabla^p R^{mn}\right) -\nabla^4R_{ij}\nb\\
&&  + \left(\nabla_i R_{mn}\right)\left(\nabla_j R^{mn}\right) +2\left(\nabla_p R_{in}\right)\left(\nabla^p R^n_j\right)\nb\\
&&  +2\nabla^n\nabla_{(i}\nabla^2R_{j)n}+2\nabla_n\left(R^n_m\nabla_{(i}R^m_{j)}\right)\nb\\
&& -2\nabla_n\left(R_{m(j}\nabla_{i)}R^{mn}\right)-2\nabla_n\left(R_{m(i}\nabla^nR^m_{j)}\right)\nb\\
 && -g_{ij}\nabla^n\nabla^m\nabla^2R_{mn}\nb\\
&=& \frac{C e^{2\nu}}{12r^{9}}\Big(180 C^2-75 C \Lambda_g r^3+189 C r \nb\\
& & ~~~~~~~~~ -4 \Lambda_g^2 r^6 \Big)\delta^{r}_{i} \delta^{r}_{j}  + \frac{C}{12r^{7}}\Big(-630 C^2 \nb\\
& & ~~~~ +150 C \Lambda_g r^3 -567 C r+2 \Lambda_g^2 r^6\Big)\Omega_{ij},
  \eqn
 where $\Omega_{ij} \equiv \delta^{\theta}_{i} \delta^{\theta}_{j} + \sin^{2}\theta\delta^{\phi}_{i} \delta^{\phi}_{j}$.

\section*{Appendix B: Slowly Rotating Vacuum Solutions}
\renewcommand{\theequation}{B.\arabic{equation}} \setcounter{equation}{0}
 Slowly rotating vacuum solutions in other versions of the HL theory
have been studied by several authors \cite{rbhs}. The goal of this section is to derive slowly rotating black hole solutions in the HMT setup. We will seek a solution of the form
\begin{eqnarray}\label{ansatz1}
 && ds^2 = - dt^2 + r^2(d\theta^2 + \sin^2 \theta d\phi^2)
  	\\ \nonumber
&&  + e^{2\nu(r)}\left[dr + e^{\mu(r) - \nu(r)}(dt - a \omega(r) \sin^2\theta d\phi)\right]^2,
\end{eqnarray}
where the functions $\nu(r), \mu(r)$, and $\omega(r)$ are independent of $(t, \theta, \phi)$. By requiring that the metric satisfy the equations to first order in the small rotation parameter $a$, we will be able to determine $\nu, \mu$, and $\omega$. 

The ansatz (\ref{ansatz1}) is motivated by the fact that it agrees with the Kerr solution to first order in $a$. Indeed, the Kerr line element expressed in Doran coordinates \cite{Doran2000} is given by
\begin{eqnarray}\label{Doran}
  ds_{\text{Kerr}}^2 &=&  -dt^2 + (r^2 + a^2 \cos^2\theta)d\theta^2 	\\ \nonumber
&&   + (r^2 + a^2)\sin^2\theta d\phi^2 + \frac{r^2 + a^2 \cos^2\theta}{r^2 + a^2}
	\\ \nonumber
&& \times \left[dr + \frac{\sqrt{2mr(r^2 + a^2)}}{r^2 + a^2 \cos^2\theta} (dt - a\sin^2\theta d\phi)\right]^2,
\end{eqnarray}
where $m$ and $a$ are parameters. As $a \to 0$, this metric coincides with (\ref{ansatz1}) to first order in the rotation parameter $a$,  provided that
$$\nu(r) = 0, \qquad \mu(r) = \log \sqrt{\frac{2m}{r}}, \qquad \omega(r) = 1.$$
In particular, when $a = 0$, it reduces to the Schwarzschild metric in Painlev\'e-Gullstrand form.

Note that the form (\ref{ansatz1}) of the line element is compatible with the projectability condition $N = N(t)$; its ADM coefficients are
$$N = 1, \qquad N^i = (e^{\mu(r) - \nu(r)}, 0, 0).$$
Working in the gauge $\varphi = 0$, the momentum constraint (\ref{eq2}) for the metric (\ref{ansatz1}) reduces to
\begin{eqnarray} \label{momentumconstraint3}
&&- \frac{2e^{\mu - 3 \nu} \nu'}{r} + O(a^2) = 0,
	\\ \nonumber
&&\frac{a e^{2\mu-2\nu}}{2 r^4} \Bigl[r^2  \left(\omega''+\omega' \left(4 \mu'-\nu'\right)\right)
	\\ \nonumber
&& -2 \omega \bigl(1-r^2  \mu''
+r^2 \mu' \nu'-2r^2  \mu'^2 +r \nu'\bigr)\Bigr]+ O(a^2) = 0,
\end{eqnarray}
while the equation (\ref{eq4b}) obtained from variation with respect to $A$ yields
\begin{equation}\label{varyAeq}  
  \left(1-r^2 \Lambda_g \right) e^{2 \nu}+2 r \nu' - 1 + O(a^2) = 0.
\end{equation}
The first equation in (\ref{momentumconstraint3}) implies that $\nu$ is constant, and then (\ref{varyAeq}) shows that
$$\nu = 0, \qquad \Lambda_g = 0.$$
This yields
\begin{eqnarray}\label{RijLKetc}
&& R_{ij} = O(a^2), 
	\\ \nonumber
&& \mathcal{L}_K = -\frac{2}{r^2}e^{2\mu}(1 + 2r\mu') + O(a^2), 
	\\ \nonumber
&& \mathcal{L}_V = 2\Lambda + O(a^2).
\end{eqnarray}

The ($rr$)-component of the dynamical equations (\ref{eq3}) gives
\bqn
\label{rrdynamical}
&& \frac{2 r A_0' - r^2\Lambda +2 r e^{2 \mu} \mu'+e^{2 \mu}}{r^2} +\frac{2 A_1'}{r} a\nb\\
& & ~~~~~~~~~~~~~~
+\mathcal{O}(a^2) = 0,  
\eqn
where we have assumed that $A(r)$ has the form
\begin{equation}\label{Aansatz}
A(r) = A_0(r) + A_1(r) a + \mathcal{O}(a^2).
\end{equation}
The terms of $\mathcal{O}(1)$ in (\ref{rrdynamical}) imply that
\begin{equation}\label{murexp}
\mu(r) = \frac{1}{2} \ln \left(\frac{2 m}{r}+\frac{1}{3} r^2 \Lambda-2
   A_0(r)+ \frac{2}{r} \int_{r_0}^{r}{A_0(s) \, ds}\right),
   \end{equation}
where $r_0 > 0$ is a constant, while the terms of $\mathcal{O}(a)$ imply that $A_1$ is a constant. With these choices, all the components of the dynamical equations as well as the equations obtained from variation with respect to $A$ and $\varphi$ are satisfied to first order in $a$, and the Hamiltonian constraint (\ref{eq1}) becomes
$$\int_0^\infty r A_0'(r) dr + \mathcal{O}(a^2) = 0.$$
Finally, the second equation in the momentum constraint (\ref{momentumconstraint3}) is satisfied to $\mathcal{O}(a)$ provided that
\bqn
\label{lambdaexpression}  
  \omega(r) &=&e^{-2\mu}\left(\frac{d_1}{r} + d_2 r^2\right)
         \\ \nonumber
& = &\frac{d_1 + d_2r^3}{2m + 2\int_{r_0}^r A_0(s) ds - 2rA_0 + \frac{\Lambda}{3}r^3}, 
 \eqn
where  $ d_1$ and $d_2$  are the integration constants.

In summary, the ansatz (\ref{ansatz1})  gives a solution to first order in $a$ provided that $\mu(r)$ is given by (\ref{murexp}), $\omega(r)$ is given by (\ref{lambdaexpression}), and
\bq\lb{Kerr}
 \nu = 0, \qquad  A(r) = A_0(r) + aA_1 + \mathcal{O}(a^2),
\eq 
where $r_0 >0, m, \Lambda, A_1, d_1, d_2$ are arbitrary constants and $A_0(r)$ can be freely chosen as long as
$$
\int_0^\infty r A_0'(r) dr = 0.
$$
We recover the slowly rotating version of the Kerr solution by taking $A_0 = 0$, $\Lambda =0$, and $d_2 = 0$. Setting $a = 0$, on the other hand, we recover the static solutions obtained
in \cite{GSW}. 

Let us point out that the standard Einstein equations also allow for a nonzero value of $d_2$ in the slowly rotating limit. Indeed, substituting the ansatz (\ref{ansatz1}) with $\nu = 0$ into the vacuum Einstein equations
$$R_{\alpha\beta} - \frac{1}{2}g_{\alpha\beta} R = 0, \qquad \alpha,\beta = 0, 1,2,3,$$
we find that they are satisfied to order $\mathcal{O}(a)$ if and only if
$$\mu(r) = \frac{1}{2}\ln \left(\frac{2m}{r}\right),$$
where $m > 0$ is a constant, and $\omega(r)$ is given by (\ref{lambdaexpression}) with arbitrary constants $d_1$ and $d_2$.


\end{document}